%% file: root.tex
\documentclass[letterpaper, 10 pt, conference]{ieeeconf}  

\IEEEoverridecommandlockouts                              

\usepackage{mathtools}
\usepackage{amsmath, amssymb}
\usepackage{empheq}
\usepackage{xcolor} 
\usepackage{nicematrix} 
\usepackage{cases} 
\usepackage{changepage} 
\usepackage{siunitx}
\usepackage{microtype} 
\usepackage{cite}

\newcounter{subfigure}[figure]

\usepackage{amsthm}

\makeatletter
\newtheoremstyle{bfnote}%
    {\baselineskip} 
    {\baselineskip} 
    {\itshape} 
    {}
    {\bfseries}
    {.}
    {.5em}
    {\thmname{#1}\thmnumber{ #2}\thmnote{ (#3)}} 

\renewenvironment{proof}[1][\proofname]{\par
    \pushQED{\qed}%
    \normalfont \topsep6\p@\@plus6\p@\relax
    \trivlist
    \item\relax
    {\normalfont\bfseries 
    #1\@addpunct{.}}\hspace\labelsep\ignorespaces
    }{%
    \popQED\endtrivlist\@endpefalse
    }
\makeatother
\theoremstyle{bfnote}
\newtheorem{theorem}{Theorem}
\newtheorem{assumption}[theorem]{Assumption}
\newtheorem{remark}[theorem]{Remark}

\usepackage{algorithm}
\usepackage{algpseudocode}

\usepackage{tikz}
\usepackage{pgfplots}
\usetikzlibrary{calc}
\usetikzlibrary {matrix}
\usetikzlibrary{shapes,arrows,patterns}
\pgfplotsset{compat=newest}
\newlength\figH
\newlength\figW
\definecolor{matlab_blue}{rgb}{0 0.4470 0.7410}
\definecolor{matlab_yellow}{rgb}{0.9290 0.6940 0.1250}
\definecolor{matlab_purple}{rgb}{0.4940 0.1840 0.5560}
\definecolor{matlab_orange}{rgb}{0.8500 0.3250 0.0980}
\definecolor{matlab_red}{rgb}{1 0 0}
\definecolor{matlab_green}{rgb}{0 1 0}

\DeclarePairedDelimiter\norm{\lVert}{\rVert}

\newcommand{\ie}{\textit{i}.\textit{e}., }
\newcommand{\eg}{\textit{e}.\textit{g}., }

\DeclareMathAlphabet{\mymathbb}{U}{bbold}{m}{n} 
\DeclareMathOperator{\eye}{\mathcal{I}}

\title{\LARGE \bf
Offset-Free Data-Driven Predictive Control: \\Implicit and Explicit Disturbance Formulations 
}

\author{Lucca Heinze Faro, Yuanbo Nie, and Paul Trodden
\thanks{This work was supported by the MSc Conversion Scholarship of the (former) Department of Automatic Control \& Systems Engineering, University of Sheffield.}
\thanks{The authors are with the School of Electrical and Electronic Engineering, University of Sheffield, S1 3JD, United Kingdom (email: lheinzefaro1@sheffield.ac.uk, y.nie@sheffield.ac.uk, p.trodden@sheffield.ac.uk}%
}

\begin{document}
\maketitle
\thispagestyle{empty}
\pagestyle{empty}

\begin{abstract}

Offset-free tracking is a key objective of any control algorithm. For predictive controllers with linear models, it is well understood that rejecting asymptotically-constant disturbances necessitates augmenting the prediction model with $p$ integrating states, where $p$ is the number of measured outputs. This can be achieved in model predictive control (MPC) by using a velocity form or a combination of disturbance model and observer. In data-driven predictive control (DPC), implicit incorporation of integrating states using the velocity form has been successfully adopted to obtain offset-free tracking; however, formulations based on an explicit disturbance model and observer have not yet been explored. In this paper, we first study how the integrating states are incorporated into the data-based representation used in DPC, translating the requirements of $p$ integrating states into the required modifications of the DPC formulation. We then propose a novel offset-free DPC formulation that incorporates an explicit constant output-disturbance model and an output disturbance observer. Numerical examples show that this formulation has similar disturbance-rejection capabilities to the velocity-form DPC, but using a smaller optimisation problem.

\end{abstract}

\section{Introduction}

Ensuring offset-free tracking whenever possible is a core goal of control algorithms \cite{pannocchia_offset-free_2005}. For general controllers, this requires that controlled variables are linearly independent and that their total number is not greater than the number of inputs or measured outputs~\cite{pannocchia_disturbance_2003}. Additionally, offset-free tracking with predictive controllers requires that~\cite{rossiter_model-based_2013}:
\begin{enumerate}
    \item in steady state, the minimum of the cost function must be consistent with zero tracking errors, and
    \item if the plant is in steady state with zero tracking error, the predicted control move must be zero.
\end{enumerate}

Condition 1) denotes having a well-posed cost function. This requires consistent steady-state input--output setpoints $(u_{}^{s},y_{}^{s})$, \ie $u_{}^{s}$ tracks $y_{}^{s}$ for the system model, as inconsistent setpoints cause a trade-off between tracking $u_{}^{s}$ and $y_{}^{s}$, which leads to offset. Common approaches to address this requirement include using consistency conditions to calculate the corresponding $u_{}^{s}$ for a given $y_{}^{s}$ \cite{pannocchia_disturbance_2003,rossiter_model-based_2013} or using an incremental-input formulation, where the input $u$ is replaced by its increment $\Delta u$ as the manipulated variable, which guarantees that $\Delta u_{}^{s}=0$ is consistent with any $y_{}^{s}$ \cite{maciejowski_implicit_1998}.

Condition 2) implies having unbiased predictions in steady state, in which case a $u_{}^{s}$ that tracks $y_{}^{s}$ for the model also tracks it for the real plant. In the presence of model mismatch, unbiased predictions can be achieved by incorporating integral action~\cite{rawlings_nonlinear_1994}. While it is a common misconception that a $\Delta u$ formulation inherently provides this capability via input integration (\ie the computation of $u$ from $\Delta u$), Maciejowski \cite{maciejowski_implicit_1998} clarifies that the differentiation in calculating $\Delta u$ cancels out the integration. In reality, integral action requires the predictive controller to incorporate a disturbance model, whose iterative estimation then eliminates steady-state offset. 

Offset-free tracking has received significant attention in the context of linear model predictive control (MPC) and is often achieved using two main approaches. The first involves explicitly augmenting the system model with integrating disturbances and using an observer to remove its effect from the controlled outputs (\eg \cite{muske_disturbance_2002,pannocchia_disturbance_2003,pannocchia_offset-free_2005}). Alternatively, integrating disturbances can be implicitly incorporated by using a velocity form, which recasts the model using state and input increments, $\Delta x$ and $\Delta u$ (\eg \cite{pannocchia_velocity_2001,betti_robust_2013}). An overview of linear MPC based on both approaches can be found \cite{pannocchia_offset-free_2015-1}.

In data-driven predictive control (DPC), the velocity form has been investigated in \cite{lazar_offsetfree_2022, hamdan_data_2024, de_jong_kernelized_2024, leng_data-driven_2026, wang_velocity-form_2026}, while an implementation based on an input disturbance model and limited to single-input systems is proposed and studied in \cite{kong_robust_2026}. Nevertheless, alternative disturbance models or a comparison between them and the velocity form have yet to be explored in the literature.

\subsection{Contribution}
The aim of this paper is to provide a systematic analysis of and comparison between the velocity form and the disturbance model in DPC. Specifically, this paper \textbf{(i)} provides a discussion on how integral action is incorporated into the velocity-form DPC, highlighting how it differs from traditional model-based predictive control, \textbf{(ii)} proposes a new offset-free DPC based on an explicit output-disturbance model, which incorporates the correct state augmentation and is applicable to multiple-input multiple-output (MIMO) systems without integrating modes, and \textbf{(iii)} compares the proposed formulation and offset-free DPCs based on the velocity form using numerical simulations.

\subsection{Outline}
   The rest of this paper is organised as follows. First, Section \ref{section:problem_statement} states the studied problem, \ie offset-free tracking under model mismatch. Section \ref{section:velocity_formulation} reviews the DPC formulation based on the velocity form. A discussion is provided on how it incorporates integral action, which shows that, unlike MPC, assuming a model with more than $p$ additional~states is not an issue. Section \ref{section:disturbance_model} proposes a novel offset-free DPC formulation, which explicitly includes an output-disturbance estimate in the optimal control problem (OCP). This controller necessitates consistency conditions and a disturbance estimator, whose formulations are also discussed. Section 
\ref{section:Formulations comparison} uses numerical examples to analyse and compare the output-disturbance DPC to the velocity-form DPC. These examples show that they have similar tracking performances, but the proposed formulation results in a smaller OCP. Section \ref{section:conclusion} provides conclusions and possible research directions.
    
\subsection{Notation}   

For a vector $z_k^{} \in \mathbb{R}^q$, $k \in \mathbb{Z}$, we denote
\begin{equation*}
    z_{[i,j]} = 
    \begin{bmatrix}
        z_{i}^\top & \ldots & z_{j}^\top
    \end{bmatrix}^\top \in \mathbb{R}^{(j-i+1)q}, \quad j \geq i, \quad i,j \in \mathbb{Z}.
\end{equation*} 
The $n$-length vector of ones is $\mymathbb{1}_n$, and the $n \times n$ identity matrix is $\eye_n$. The Kronecker product is denoted by $\otimes$. 

The Hankel matrix of ${z}_{[k-T+1,k]}$, $z_i \in \mathbb{R}^q$, is given by
\begin{equation}
    \label{eq:Hankel_matrix}
    \mathcal{H}_{L}^{} (z_{[k-T+1,k]}^{})
    \coloneqq \begin{bmatrix}
        z_{k-T+1}^{}& \cdots & z_{k-L+1}^{} \\
        \vdots &  & \vdots \\
        z_{k-T+L}^{}& \cdots & z_{k}^{}
    \end{bmatrix}.
\end{equation}
The signal ${z}_{[k-T+1,k]}^{}$ is said to be \emph{persistently exciting} (PE) of order $L$ if its Hankel matrix $\mathcal{H}_{L}^{} (z_{[k-T+1,k]}^{})$ has rank $q L$.

\input{2-Problem_Statement}

\section{Conclusions}
\label{section:conclusion}

This paper studied the use of offset-free formulations in DPC and its contributions are \textbf{(i)} a discussion of how integral action is incorporated into the data-driven representation and \textbf{(ii)} a novel offset-free DPC formulation based on an output disturbance model, which does not require trajectory augmentation. The numerical simulations show that this proposed formulation can obtain comparable tracking performance to the velocity-form DPC whilst reducing computational complexity, even when the underlying plant is nonlinear. 

The proposed formulation adopted an output-disturbance model because of its simple implementation, but this choice limits the application of the output-disturbance DPC to system without integrating modes. Future work should look at how more general disturbance models can be explicitly incorporated into DPC to avoid this limitation.



\bibliographystyle{IEEEtran}
\bibliography{PhD}



\addtolength{\textheight}{-12cm}   


\end{document}

%% file: 2-Problem_Statement.tex

\section{Problem Statement}
\label{section:problem_statement}

\begin{subequations} \label{eq:data_driven_MPC_entire_formulation}
    Consider the DPC controller adapted from~\cite{berberich_data-driven_2021-1}:
    \begin{alignat}{3}
        & \min_{\substack{ \alpha, \, \Bar{u}_{[-n,N-1]}^{}, \\ \sigma,  \,\Bar{y}_{[-n,N-1]}^{}}} & & & \begin{split} \sum_{i=0}^{N-1} \norm{\Bar{u}_{i}^{} - u_{}^{s}}_{R}^{2} + &\norm{\Bar{y}_{i}^{} - y_{}^{s}}_Q^{2}  \\[-5pt] + \lambda_{\alpha} &\norm{\alpha}_2^2 + \lambda_{\sigma} \norm{\sigma}_2^2 \end{split} \label{eq:data_driven_MPC_cost}\\
        & & \llap{\text{s.t.}} & &
        \begin{bNiceMatrix}
            \Bar{u}_{[-n,N-1]}^{} \\ \Bar{y}_{[-n,N-1]}^{} + \sigma
        \end{bNiceMatrix}
        &= 
        \begin{bNiceMatrix}
           \mathcal{H}_{\scriptscriptstyle N+n}^{} ({u}^{d})  \\ 
           \mathcal{H}_{\scriptscriptstyle N+n}^{} ({y}^{d})
        \end{bNiceMatrix} \alpha, \label{eq:data_driven_MPC_dynamics} \\
        & & & & \begin{bNiceMatrix}
            \Bar{u}_{[-n,-1]}^{} \\ \Bar{y}_{[-n,-1]}^{}
        \end{bNiceMatrix}
        & = 
        \begin{bNiceMatrix}
           {u}_{[k-n,k-1]}^{}  \\ {y}_{[k-n,k-1]}^{}
        \end{bNiceMatrix}, \label{eq:data_driven_MPC_initial}\\
        & & & & \begin{bNiceMatrix}
            \Bar{u}_{[N-n,N-1]}^{} \\ \Bar{y}_{[N-n,N-1]}^{}
        \end{bNiceMatrix}
        & = 
        \begin{bNiceMatrix}
           \mymathbb{1}_n \otimes {u}_{}^{s}  \\ \mymathbb{1}_n \otimes {y}_{}^{s}
        \end{bNiceMatrix}, \label{eq:data_driven_MPC_terminal} \\
        & & & & \Bar{u}_{i}^{} \in \mathbb{U}, & \quad i = 0, \ldots,N-1,
        \label{eq:data_driven_MPC_inputoutput}
    \end{alignat}
    where $(\Bar{u}_{[-n,N-1]}^{}, \Bar{y}_{[-n,N-1]}^{}) \in \mathbb{R}^{(N+n)m} \times \mathbb{R}^{(N+n)p} $ are predicted input--output trajectories, $\sigma \in \mathbb{R}^{(N+n)p}$ is a slack, $(u_{}^{d},y_{}^{d}) \in \mathbb{R}^{mT} \times \mathbb{R}^{pT}$ are input--output trajectories data, and $\alpha \in \mathbb{R}^{T-(N+n)+1}$ is a decision variable. The input applied at time $k$ is $u_k^{} = \Bar{u}_{0}^{*}$, where $\Bar{u}_{[-n,N-1]}^{*}$ is the solution to~\eqref{eq:data_driven_MPC_entire_formulation}.
\end{subequations}

The DPC formulation is based on Willems' Fundamental Lemma \cite{willems_note_2005}, which defines sufficient conditions for data-driven trajectory representation. As trajectories in \eqref{eq:data_driven_MPC_dynamics} have length $N+n$, these conditions are satisfied by Assumption~\ref{assumption:7-initial_data}.

\begin{assumption} \label{assumption:7-initial_data}
    The input trajectory $u_{}^{d}$ is persistently exciting of order $L=N+2n$.
\end{assumption}

This paper investigates the problem of achieving offset-free output tracking with the DPC \eqref{eq:data_driven_MPC_entire_formulation} in the presence of model mismatch, more specifically when $(u_{}^{d},y_{}^{d})$ is a trajectory of a minimal LTI model with $(u_k^{}, y_k^{}, x_k^{}) \in \mathbb{R}^m \times \mathbb{R}^p \times \mathbb{R}^n$,%

\begin{subequations} \label{eq:linear_dynamics2}%
    \leavevmode\vspace{-\dimexpr\topskip \relax}
    \begin{empheq}[left={\mathcal{M}: \quad \empheqlbrace}]{align}
        x_{k+1} &= A x_{k}^{} + B u_{k}^{},\\
        y_{k} &= C x_{k}^{},
    \end{empheq}
\end{subequations}%
\begin{subequations} \label{eq:plant}%
    which is different from the underlying controlled plant,
    \vspace{-\abovedisplayskip}
    \begin{empheq}[left={\mathcal{P}: \quad \empheqlbrace}]{align}
        x_{k+1} &= A x_{k}^{} + B u_{k}^{} + d_x,\\
        y_{k} &= C x_{k}^{} + d_y,
    \end{empheq}
    subject to state and output disturbances, $d_x$ and $d_y$. In this case, the setpoints $(u_{}^{s}, y_{}^{s})$ may not be consistent and the data-driven representation \eqref{eq:data_driven_MPC_dynamics} may not provide unbiased predictions in steady state, which can lead to steady-state offset with the DPC~\eqref{eq:data_driven_MPC_entire_formulation}. 
\end{subequations}
Assumption~\ref{assumption:existence_input_setpoint} ensures that steady-state disturbances can be compensated \cite{pannocchia_disturbance_2003} and thus an offset-free formulation can be devised. 

\begin{assumption} \label{assumption:existence_input_setpoint}
    The matrix $\begin{bmatrix} A - \eye_n & B \\ C & 0 \end{bmatrix}$ has full row rank.
\end{assumption}


\section{Implicit Disturbance Model in DPC}
\label{section:velocity_formulation}

This section discusses the velocity form in DPC for offset-free tracking. First, a DPC based on the velocity form, as used in \cite{lazar_offsetfree_2022,de_jong_kernelized_2024, hamdan_data_2024,leng_data-driven_2026}, is reviewed, highlighting the required adaptations to the standard DPC formulation \eqref{eq:data_driven_MPC_entire_formulation}. Then, a discussion is provided on how these adaptations implicitly incorporate integral action into DPC.

\subsection{Review of Velocity-form DPC}
\label{subsection: Review of velocity-form DPC}

\begin{subequations} \label{eq:augmented_system}
    \setlength{\abovedisplayskip}{5pt}
    \setlength{\belowdisplayskip}{5pt}
    Offset-free linear MPC requires dynamics augmentation with $n_d = p$ integrating disturbances \cite[Lemma 1.10]{rawlings_model_2020}, where $p$ is the number of measured outputs. This~is because $n_d \leq p$ is necessary for detectability of the augmented system (which ensures that the disturbances can be observed \cite{pannocchia_disturbance_2003}), while $n_d \geq p$ is necessary to guarantee that convergence of the disturbance estimate corresponds to zero offset. This condition is satisfied by the augmented model 
    \begin{align}
        \begin{bmatrix}
            \Bar{x}_{k+1}^{} \\ \Bar{\delta}_{k+1}^{}
        \end{bmatrix} &= 
        \begin{bmatrix}
            A & B_d \\ \mymathbb{0} & \eye_p
        \end{bmatrix}
        \begin{bmatrix}
            \Bar{x}_{k}^{} \\ \Bar{\delta}_{k}^{}
        \end{bmatrix} + 
        \begin{bmatrix}
            B \\ \mymathbb{0}
        \end{bmatrix} \Bar{u}_{k}^{}, \\
        \Bar{y}_{k}^{} &=
        \begin{bmatrix}
            C & C_d
        \end{bmatrix}
        \begin{bmatrix}
            \Bar{x}_{k}^{} \\ \Bar{\delta}_{k}^{}
        \end{bmatrix},
    \end{align} 
\end{subequations}
\begin{subequations}  \label{eq:6-velocity_formulation}
    \setlength{\abovedisplayskip}{5pt}
    \setlength{\belowdisplayskip}{5pt}
    which can be recast as the velocity form of \eqref{eq:linear_dynamics2}
    \begin{align}
        \begin{bmatrix}
            \Delta \Bar{x}_{k+1}^{} \\ \Bar{z}_{k+1}^{}
        \end{bmatrix} &= 
        \begin{bmatrix}
            A & \mymathbb{0 } \\ CA & \eye_p
        \end{bmatrix}
        \begin{bmatrix}
            \Delta \Bar{x}_{k}^{} \\ \Bar{z}_{k}^{}
        \end{bmatrix} + 
        \begin{bmatrix}
            B \\ CB
        \end{bmatrix}
        \Delta \Bar{u}_{k}^{}, \\
        \Bar{y}_{k}^{} &=
        \begin{bmatrix}
            \mymathbb{0} & \eye_p
        \end{bmatrix}
        \begin{bmatrix}
            \Delta \Bar{x}_{k}^{} \\ \Bar{z}_{k}^{}
        \end{bmatrix},
    \end{align}
    using input and state increments $(\Delta \bar{u}, \Delta \bar{x})$.
\end{subequations}

\begin{remark}
    It follows from controllability and observability of \eqref{eq:linear_dynamics2} and from Assumption \ref{assumption:existence_input_setpoint} that the velocity form is also controllable and observable. This can be verified using Hautus lemmas for controllability and observability \cite{rawlings_model_2020}.
\end{remark}

The velocity form \eqref{eq:6-velocity_formulation} can be incorporated into DPC \eqref{eq:data_driven_MPC_entire_formulation} by replacing input variables and parameters by their incremental counterparts, \ie $(u,\Bar{u},u_{}^{s}, \mathbb{U}_k) \rightarrow (\Delta u, \Delta \Bar{u}, 0, \Delta \mathbb{U}_k)$, and increasing the model order from $n$ to $n+p$, resulting in:%

\begin{subequations} \label{eq:7-DPC_entire_formulation_velocity} %
    \leavevmode\vspace{-\dimexpr\topskip \relax}
    \begin{alignat}{3}
        & \min_{\substack{ \alpha, \, \Bar{u}, \\ \sigma,  \,\Bar{y}}} \;& & \sum_{i=0}^{N-1} \norm{\Delta  \Bar{u}_{i}^{}}_{R}^{2} + \norm{\Bar{y}_{i}^{} - y_{}^{s}}_Q^{2} \lambda_{\alpha} \norm{\alpha}_2^2 + \lambda_{\sigma} \norm{\sigma}_2^2 \label{eq:7-DPC_cost_velocity}\\
        & & \llap{\text{s.t.}} \;&
        \begin{bNiceMatrix}
            \Delta \Bar{u}_{[-n-p,N-1]}^{} \\[4pt] 
            \Bar{y}_{[-n-p,N-1]}^{} + \sigma
        \end{bNiceMatrix}=
        \begin{bNiceMatrix}
           \mathcal{H}_{N+n+p}^{} (\Delta {u}_{}^{d})  \\[4pt]
           \mathcal{H}_{N+n+p}^{} ({y}^{d})
        \end{bNiceMatrix} \alpha, \label{eq:7-DPC_dynamics_velocity} \\[0pt]
        & && \begin{bmatrix}
            \Delta \Bar{u}_{[-n-p,-1]}^{} \\[4pt] 
            \Bar{y}_{[-n-p,-1]}^{}
        \end{bmatrix} = 
        \begin{bmatrix}
           \Delta {u}_{[-n-p,-1]}^{}  \\[4pt] 
           {y}_{[-n-p,-1]}^{}
        \end{bmatrix},  \label{eq:7-DPC_initial_velocity} \\[0pt]
        & &&  \begin{bmatrix}
            \Delta \Bar{u}_{[N-n-p,N-1]}^{} \\[4pt] 
            \Bar{y}_{[N-n-p,N-1]}^{}
        \end{bmatrix} = 
        \begin{bmatrix}
           \mymathbb{0}  \\[4pt] 
           \mymathbb{1}_{n+p} \otimes {y}_{}^{s}
        \end{bmatrix}, \label{eq:7-DPC_terminal_velocity} \\[0pt]
        & && \Delta \Bar{u} \in \Delta \mathbb{U}_k,   \label{eq:7-DPC_inputoutput_velocity}
    \end{alignat}
\end{subequations}  
termed velocity-form DPC. The set $\Delta \mathbb{U}_k$ must ensure that $\Delta \Bar{u} \in \Delta \mathbb{U}_k \Longrightarrow \Bar{u} \in \mathbb{U}^{N+n+p}$.

\begin{remark}
    The velocity form \eqref{eq:6-velocity_formulation} embeds the disturbance model in the calculation of the initial state $(\Delta \bar{x}_0^{},\Delta \bar{z}_0^{})$~\cite{pannocchia_velocity_2001}. Consequently, the velocity-form DPC \eqref{eq:7-DPC_entire_formulation_velocity} implicitly incorporates the disturbance via the initial equality constraint~\eqref{eq:7-DPC_initial_velocity}.
\end{remark}

\begin{remark}
    The use of the augmented model \eqref{eq:6-velocity_formulation} corresponds to an augmented predicted trajectory in \eqref{eq:7-DPC_dynamics_velocity} compared to \eqref{eq:data_driven_MPC_dynamics}, increasing the number of variables and constraints.
\end{remark}

\subsection{Effects of the model order on the velocity-form DPC }

As discussed in Section \ref{subsection: Review of velocity-form DPC}, the requirement for $n_d=p$ disturbances in offset-free linear MPC comes partially from detectability of the augmented model, which requires that $n_d \leq p$. If more disturbances are used, \ie $n_d > p$, a stable estimator cannot be designed, resulting in unstable and unobserved disturbances in the dynamics \cite{pannocchia_disturbance_2003}. We argue that, contrary to MPC, this is not an issue in DPC.

To explain why, we recall that the data-driven representation enabled by Willems' Fundamental Lemma \cite{willems_note_2005} describes an input--output behaviour, not a specific state realisation as in MPC. The properties of the data-driven representation are those of a \emph{minimal} realisation, which describes the input--output dynamics with the fewest states possible~\cite{markovsky_data-driven_2023}.

\begin{subequations} \label{eq:6-incremental_formulation}
    Alternative augmented $(\Delta u, y)$ models with more than $n+p$ states, such as the incremental-input formulation with overactuated systems (\ie $m>p$),
    \begin{align}
        \begin{bmatrix}
            \Bar{x}_{k+1}^{} \\ \Bar{u}_{k}^{}
        \end{bmatrix} &= 
        \begin{bmatrix}
            A & B \\ \mymathbb{0} & \eye_m
        \end{bmatrix}
        \begin{bmatrix}
            \Bar{x}_{k}^{} \\ \Bar{u}_{k-1}^{}
        \end{bmatrix} + 
        \begin{bmatrix}
            B \\ \eye_m
        \end{bmatrix} \Delta \Bar{u}_{k}^{}, \\
        \Bar{y}_{k}^{} &=
        \begin{bmatrix}
            C & \mymathbb{0}
        \end{bmatrix}
        \begin{bmatrix}
            \Bar{x}_{k}^{} \\ \Bar{u}_{k-1
            }^{}
        \end{bmatrix},
    \end{align}  %
\end{subequations} %
\begin{subequations} \label{eq:7-augmented_velocity}%
    or the velocity form with extra $v \in \mathbb{R}^{n_v}$ disturbances, $n_v \geq 1$,
    \begin{align} 
        \begin{bmatrix}
            \Delta \Bar{x}_{k+1}^{} \\ z_{k+1}^{} \\ v_{k+1}^{}
        \end{bmatrix} &= 
        \begin{bmatrix}
            A & \mymathbb{0} & \mymathbb{0} \\ CA & \eye_p & \mymathbb{0} \\ \mymathbb{0} & \mymathbb{0} & \eye_{n_v}
        \end{bmatrix}
        \begin{bmatrix}
            \Delta \Bar{x}_{k}^{} \\ z_{k}^{} \\ v_{k}^{}
        \end{bmatrix} + 
        \begin{bmatrix}
            B \\ CB \\ B_v
        \end{bmatrix}
        \Delta \Bar{u}_{k}^{}, \\
        \Bar{y}_{k}^{} &=
        \begin{bmatrix}
            \mymathbb{0} & \eye_p & \mymathbb{0}
        \end{bmatrix}
        \begin{bmatrix}
            \Delta \Bar{x}_{k}^{} \\ z_{k}^{} \\ v_{k}^{}
        \end{bmatrix},
    \end{align}
    describe the same input--output dynamics as the velocity form~\eqref{eq:6-velocity_formulation}, thus leading to identical data-driven representations. They are, however, not observable and nonminimal, and the velocity form is a minimal realisation. Thus, it follows that for a DPC using $(\Delta u, y)$ and a model order of $n+p$ or more, the data-driven representation corresponds to the velocity form~\eqref{eq:6-velocity_formulation} (or any equivalent minimal model), which implicitly incorporates the correct number of disturbances, \ie $p$.  
\end{subequations}

Consequently, nonminimal models lead to DPC formulations nearly identical to \eqref{eq:7-DPC_entire_formulation_velocity}, the only difference being the~order of the augmented model, which overestimates that of the data-driven representation, \ie $n+p$. However, explicit uses of the model order in DPC (\ie the boundary conditions length and the persistence of excitation depth) only require an upper bound on the model order, not the exact value~\cite{berberich_data-driven_2021-1}, which is satisfied by these overestimations. 

\begin{remark}
    The consequence of designing a DPC controller based on a $(\Delta u, y)$ model with more than $n+p$ states is an increase in the number of variables and constraints in the DPC formulation, unnecessarily adding complexity to the resulting OCP. Thus, an augmented order of $n+p$ is preferred, as it correctly incorporates the velocity form while resulting in a smaller number of variables and constraints.
\end{remark}


\section{Explicit Disturbance Model in DPC}
\label{section:disturbance_model}

\begin{subequations} \label{eq:7-LTI_constant_disturbance}
    \setlength{\abovedisplayskip}{5pt}
    \setlength{\belowdisplayskip}{5pt}
    This section proposes a novel offset-free DPC formulation inspired by the constant output-disturbance model~\cite{rawlings_nonlinear_1994}
    \begin{align} 
        \begin{bmatrix}
            \Bar{x}_{k+1}^{} \\ \Bar{\delta}_{k+1}^{}
        \end{bmatrix} &= 
        \begin{bmatrix}
            A & \mymathbb{0} \\ \mymathbb{0} & \eye_p
        \end{bmatrix}
        \begin{bmatrix}
            \Bar{x}_{k}^{} \\ \Bar{\delta}_{k}^{}
        \end{bmatrix} + 
        \begin{bmatrix}
            B \\ \mymathbb{0}
        \end{bmatrix}
        \Bar{u}_{k}^{}, \\
        \Bar{y}_{k}^{} &= 
        \begin{bmatrix}
            C & \eye_p
        \end{bmatrix}
        \begin{bmatrix}
            \Bar{x}_{k}^{} \\ \Bar{\delta}_{k}^{}
        \end{bmatrix},
    \end{align}
    which corresponds to the augmented model \eqref{eq:augmented_system} with $B_d = \mymathbb{0}$ and $C_d = \eye_p$. The proposed formulation differs from the velocity-form DPC \eqref{eq:7-DPC_entire_formulation_velocity} in that it \emph{explicitly} includes $p$ output disturbances. Moreover, unlike the offset-free DPC in \cite{kong_robust_2026}, which uses an input disturbance and is limited to single-input systems, the proposed DPC works with MIMO systems.   
\end{subequations}

Choosing a constant output disturbance enables its easy incorporation into DPC and avoids the trajectory augmentation required by the velocity-form DPC~\eqref{eq:7-DPC_entire_formulation_velocity}. However, it requires that the unaugmented model \eqref{eq:linear_dynamics2}, and thus the corresponding $(u_{}^{d},y_{}^{d})$, does not contain integrating modes, \ie Assumption \ref{assumption:no_integrators} must hold. This is required as the disturbance cannot be distinguished from inherent integrating modes of the plant~\cite{pannocchia_disturbance_2003}.

\begin{assumption} \label{assumption:no_integrators}
    Matrix $A$ has no eigenvalues at $z=1$.
\end{assumption}
\begin{remark}
    From Assumption \ref{assumption:no_integrators} and observability of \eqref{eq:linear_dynamics2}, it follows that the augmented system \eqref{eq:7-LTI_constant_disturbance} is observable \cite{rawlings_nonlinear_1994}.
\end{remark}

\subsection{Output-Disturbance DPC}
   
Although $\Bar{\delta}_k^{}$ is not controllable, observability of \eqref{eq:7-LTI_constant_disturbance} allows the use of a disturbance observer to eliminate its effect from the dynamics \cite{pannocchia_disturbance_2003}. This can be achieved by noting that 
\begin{equation*}
    \setlength{\abovedisplayskip}{2pt}
    \setlength{\belowdisplayskip}{3pt}
    \Bar{\delta}_{k+1}^{} = \Bar{\delta}_{k}^{}
\end{equation*}
is a constant in the prediction model \eqref{eq:7-LTI_constant_disturbance},%
\begin{subequations} \label{eq:LTI_constant_disturbance_reformulation2}
    \setlength{\abovedisplayskip}{5pt}%
    \setlength{\belowdisplayskip}{0pt}%
    and the disturbance dynamics is entirely assigned to the estimator. The augmented model \eqref{eq:7-LTI_constant_disturbance} can thus be simplified by defining $\widehat{\delta} = \delta_{k+1}^{} = \delta_{k}^{}$ as the current disturbance estimate and recasting \eqref{eq:7-LTI_constant_disturbance} as
    \begin{align}
        \Bar{x}_{k+1}^{} &= A \Bar{x}_{k}^{} + B \Bar{u}_{k}^{}, \\
        \Bar{y}_{k}^{} &= C \Bar{x}_{k}^{} + \widehat{\delta}.
    \end{align}
\end{subequations}

\begin{remark}
    As $\widehat{\delta}$ is estimated at each time step, it would~be more accurately represented as $\widehat{\delta}_{k}^{}$. However, as it is constant over the prediction horizon, while other variables are time varying, the notation $\widehat{\delta}$ is adopted in \eqref{eq:LTI_constant_disturbance_reformulation2} to avoid confusion.
\end{remark}
    
\begin{remark}
    While estimating $\widehat{\delta}$ requires an observer on the augmented system \eqref{eq:7-LTI_constant_disturbance}, treating $\widehat{\delta}$ as a constant offset in the prediction model \eqref{eq:LTI_constant_disturbance_reformulation2} prevents DPC from carrying additional states, keeping the model complexity at $n$ rather than $n+p$.
\end{remark} 

A data-driven representation of the output-disturbance model \eqref{eq:LTI_constant_disturbance_reformulation2} can be obtained by defining the change of variable
\begin{equation} \label{eq:7-change_variables}
    \setlength{\abovedisplayskip}{3pt}
    \setlength{\belowdisplayskip}{3pt}
    \widehat{y}_{k}^{} = \Bar{y}_{k}^{} - \widehat{\delta},
\end{equation}
and noticing that applying it to \eqref{eq:LTI_constant_disturbance_reformulation2} results in the same input--output relation as the nominal model \eqref{eq:linear_dynamics2}. Consequently,%
\begin{equation*}
    \setlength{\abovedisplayskip}{5pt}
    \setlength{\belowdisplayskip}{5pt}
    \begin{bmatrix}
        \Bar{u}_{[-n,N-1]}^{} \\[4pt] 
        \widehat{y}_{[-n,N-1]}^{}
    \end{bmatrix} =
    \begin{bmatrix}
       \mathcal{H}_{N+n}^{} ({u}_{}^{d})  \\[4pt]
       \mathcal{H}_{N+n}^{} ({y}_{}^{d})
    \end{bmatrix} \alpha
    \; \Longleftrightarrow \,
    \begin{aligned}
        \Bar{x}_{k+1}^{} &= A \Bar{x}_{k}^{} + B \Bar{u}_{k}^{}, \\
        \widehat{y}_{k}^{} &= C \Bar{x}_{k}^{},
    \end{aligned}
\end{equation*}
\ie the data $(u_{}^{d},y_{}^{d})$ can represent trajectories of $(\Bar{u},\widehat{y})$.

\begin{subequations} \label{eq:7-DPC_disturbance_mode_robust} 
    \setlength{\abovedisplayskip}{5pt}
    \setlength{\belowdisplayskip}{5pt}
    From this observation, the output disturbance model can be incorporated into DPC by adapting \eqref{eq:data_driven_MPC_entire_formulation} to track $\widehat{y}$ instead of $\Bar{y}$, leading to the output-disturbance DPC
    \begin{alignat}{3}
        & \min_{\substack{\alpha, \, \sigma, \\ \Bar{u}, \, \widehat{y}}}  && \begin{aligned} 
            \sum_{i=0}^{N-1} \norm{\Bar{u}_{i}^{} - u_{k}^{s}}_{R}^{2} + \norm{\widehat{y}_{i}^{} &- y_{}^{s} + \widehat{\delta}}_Q^{2}  \\[-7pt]
            &+ \lambda_{\alpha} \norm{\alpha}_2^2 + \lambda_{\sigma} \norm{\sigma}_2^2 \end{aligned} \label{eq:7-DPC_disturbance_mode_cost_robust}\\
        & \;\; \quad \text{s.t.} &&
        \begin{bmatrix}
            \Bar{u}_{[-n,N-1]}^{} \\[4pt] 
            \widehat{y}_{[-n,N-1]}^{} + \sigma
        \end{bmatrix} =
        \begin{bmatrix}
           \mathcal{H}_{N+n}^{} ({u}_{}^{d})  \\[4pt]
           \mathcal{H}_{N+n}^{} ({y}_{}^{d})
        \end{bmatrix} \alpha, \label{eq:7-DPC_disturbance_mode_dynamics_robust} \\[0pt]
        & && \begin{bmatrix}
            \Bar{u}_{[-n,-1]}^{} \\[4pt] 
            \widehat{y}_{[-n,-1]}^{}
        \end{bmatrix} = 
        \begin{bmatrix}
           {u}_{[k-n,k-1]}^{}  \\[4pt] 
           {y}_{[k-n,k-1]}^{} - \widehat{\delta}
        \end{bmatrix},  \label{eq:7-DPC_disturbance_mode_initial_robust} \\[0pt]
        & &&  \begin{bmatrix}
            \Bar{u}_{[N-n,N-1]}^{} \\[4pt] 
            \widehat{y}_{[N-n,N-1]}^{}
        \end{bmatrix} = 
        \begin{bmatrix}
           \mymathbb{1}_{n} \otimes u_{k}^{s}  \\[4pt] 
           \mymathbb{1}_{n} \otimes (y_{}^{s} - \widehat{\delta})
        \end{bmatrix}, \label{eq:7-DPC_disturbance_mode_terminal_robust} \\[0pt]
        & && \Bar{u}_{i}^{} \in \mathbb{U},  \quad i = 0, \ldots,N-1. \label{eq:7-DPC_disturbance_mode_inputoutput_robust}
    \end{alignat} 
    The adaptations are based on the change of variables \eqref{eq:7-change_variables} and involve replacing $(y,\Bar{y},y_{}^{s}) \rightarrow (y-\widehat{\delta}, \widehat{y}, y_{}^{s} - \widehat{\delta})$. This incorporates the disturbance $\widehat{\delta}$ into the dynamics, the initial and terminal constraints, and the cost function. As the value of $\widehat{\delta}$ is assumed to be known when solving~\eqref{eq:7-DPC_disturbance_mode_robust}, its inclusion does not increase the problem size. 
\end{subequations}

\subsection{Consistency Conditions}
Due to the disturbance, the setpoints $(u_{}^{s},y_{}^{s}-\widehat{\delta})$ are not consistent and can cause offset. This is addressed in the output-disturbance DPC \eqref{eq:7-DPC_disturbance_mode_robust} by adopting a varying input setpoint $u_{k}^{s}$, determined such that $ \left( \mymathbb{1}_{N+n}^{} \otimes u_{k}^{s}, \; \mymathbb{1}_{N+n}^{} \otimes (y_{}^{s} - \widehat{\delta}) \right)$ is a trajectory of the prediction model \eqref{eq:LTI_constant_disturbance_reformulation2}.

Data-driven consistency conditions can be derived by noting that, in steady state, the data-driven dynamics are given by 
\begin{equation*}
    \setlength{\abovedisplayskip}{3pt}%
    \setlength{\belowdisplayskip}{3pt}%
    \begin{bmatrix}
        \mymathbb{1}_{n}^{} \otimes u_{k}^{s} \\[4pt] 
        \mymathbb{1}_{n}^{} \otimes (y_{}^{s} - \widehat{\delta})
    \end{bmatrix} =
    \begin{bmatrix}
       \mathcal{H}_{n}^{} ({u}_{}^{d})  \\[4pt]
       \mathcal{H}_{n}^{} ({y}_{}^{d})
    \end{bmatrix} \alpha_{}^s,
\end{equation*}
which can be recast as
\begin{equation} \label{eq:7-consistency_conditions}
    \setlength{\abovedisplayskip}{5pt}%
    \setlength{\belowdisplayskip}{5pt}%
    \begin{bmatrix}
        \mymathbb{1}_{n}^{} \otimes \eye_{m}^{}  & -\mathcal{H}_{n}^{} ({u}_{}^{d}) \\ 
        \mymathbb{0} & -\mathcal{H}_{n}^{} ({y}_{}^{d})
    \end{bmatrix} 
    \begin{bmatrix}
        u_{k}^{s} \\ \alpha_{}^s
    \end{bmatrix}= 
    \begin{bmatrix}
         \mymathbb{0} \\
        - \mymathbb{1}_{L}^{} \otimes (y_{}^{s} - \widehat{\delta})
    \end{bmatrix}.
\end{equation}

System \eqref{eq:7-consistency_conditions} has $n(m+p)$ equations and $T-n+1 +m$ variables and is overdetermined for the minimum $T$ satisfying the PE requirement for \eqref{eq:7-DPC_disturbance_mode_robust}. Nevertheless, existence of a solution is ensured by Assumption \ref{assumption:existence_input_setpoint}. This can be verified by studying the equivalent model-based consistency equation
\begin{equation*}
    \setlength{\abovedisplayskip}{3pt}%
    \setlength{\belowdisplayskip}{3pt}%
    \begin{bmatrix}
        A-\eye_n & B \\ C & \mymathbb{0}
    \end{bmatrix}
    \begin{bmatrix}
        x_{k}^{s} \\ u_{k}^{s}
    \end{bmatrix} = 
    \begin{bmatrix}
        \mymathbb{0} \\ y_{}^{s} - \widehat{\delta}
    \end{bmatrix},
\end{equation*}   
and noting that it has a solution  if $\left[ \begin{smallmatrix} A - \eye_n & B \\ C & 0 \end{smallmatrix} \right]$ has full row rank, \ie if Assumption \ref{assumption:existence_input_setpoint} holds. Furthermore, this solution is unique if $\left[ \begin{smallmatrix} A - \eye_n & B \\ C & 0 \end{smallmatrix} \right]$ is also square, which requires $m=p$.

\begin{remark}
    \setlength{\abovedisplayskip}{5pt}
    \setlength{\belowdisplayskip}{5pt}
    The consistency condition \eqref{eq:7-consistency_conditions} implicitly assumes that the setpoint $(u_{k}^{s}, y_{}^{s})$ is admissible, \ie $u_{k}^{s} \in \mathbb{U}$. If this is not true, or if $u_{k}^{s}$ is not unique, the linear system~\eqref{eq:7-consistency_conditions} can be replaced by a setpoint optimisation \cite{berberich_data-driven_2020}
    \begin{subequations}
        \begin{alignat}{3}
            & \min_{\substack{ \alpha, \, u_{k}^{s}, \, y_{k}^{s}}}  \quad && \norm{y_{k}^{s} - y_{}^{s}}_Q^{2} + \norm{u_{k}^{s} - u_{}^{s}}_R^{2}\\
            & \;\; \quad \text{s.t.} &&
            \begin{bmatrix}
                \left(\mymathbb{1}_{L}^{} \otimes \eye_{m}^{} \right) u_{k}^{s} \\[4pt] 
                \left(\mymathbb{1}_{L}^{} \otimes \eye_{p}^{} \right) (y_{k}^{s} - \widehat{\delta})
            \end{bmatrix} =
            \begin{bmatrix}
               \mathcal{H}_{L}^{} ({u}_{}^{d})  \\[4pt]
               \mathcal{H}_{L}^{} ({y}_{}^{d})
            \end{bmatrix} \alpha, \\
            & && u_{k}^{s} \in \mathbb{U},
        \end{alignat} 
    \end{subequations}
    where $(u_{k}^{s},y_{k}^{s})$ are optimal reachable setpoints. This setpoint optimisation also yields a time-dependent output setpoint $y_{k}^{s}$, which should replace the fixed $y_{}^{s}$ in \eqref{eq:7-DPC_disturbance_mode_robust}.
\end{remark}

\subsection{Disturbance Estimation}

The decoupling between disturbance estimation and trajectory prediction allows the disturbance estimation to be incorporated into DPC without a trajectory augmentation, but necessitates the use of an observer.%

As only an estimate of the output disturbance is required, rather than the full state as in MPC, the disturbance observer 
\begin{equation} \label{eq:deadbeat_observer}
    \setlength{\abovedisplayskip}{5pt}
    \setlength{\belowdisplayskip}{5pt}
    \widehat{\delta}_{k+1}^{} = \widehat{\delta}_{k}^{} + \gamma (y_{k}^{} - \bar{y}_{k}^{}), \quad \gamma \leq 1
\end{equation}
is adopted. This observer only uses output predictions and measurements and does not require any model knowledge, which may not be available in DPC applications. Moreover, the parameter $\gamma$ provides low-pass filtering action for $\gamma<1$, which can improve performance in the presence of noise. 

At each time step $k$, the latest disturbance estimate $\widehat{\delta}_k^{}$ is substituted into the DPC~\eqref{eq:7-DPC_disturbance_mode_robust} as $\widehat{\delta}$.

\subsection{Trade-offs with the velocity-form DPC}

The proposed solution introduces some trade-off when compared to the velocity-form DPC. First, it removes the requirement for trajectory augmentation, maintaining the complexity of the standard DPC \eqref{eq:data_driven_MPC_entire_formulation}, but it necessitates the use of consistency conditions and a disturbance estimator.

Assuming each formulation uses the shortest data length~$T$ that ensures their respective persistence of excitation requirements, the velocity-form DPC \eqref{eq:7-DPC_entire_formulation_velocity} results in an OCP with
{
    \setlength{\abovedisplayskip}{2pt}
    \setlength{\belowdisplayskip}{3pt}
    \begin{align*}
        & (N+n+p)(2m + 2p) + n + p \text{ variables and} \\
        & (N+3n+3p)(m+p) \text{ linear equality constraints}.
    \end{align*}
}%
This corresponds to additional $2p^2 + p(2m+1)$ variables and $3p^2+3mp$ constraints, compared to the output-disturbance DPC. This difference in problem size does not depend~on the prediction horizon $N$, but grows quadratically with~the number of outputs $p$ and linearly with the number of~inputs~$m$. Conversely, the output-disturbance DPC requires the consistency condition \eqref{eq:7-consistency_conditions}, which has $n(m+p)$ equations and $(m+1)(N+m)+m$ variables. The output disturbance observer~\eqref{eq:deadbeat_observer} only requires simple operations with available quantities, and thus its computational cost is negligible.

In addition to OCP complexity, the proposed formulation also introduces a trade-off in terms of controller applicability. While the output-disturbance DPC is limited to systems without integrating modes (a standard necessary condition to ensure observability in output disturbance-based approaches to tracking), the velocity-form DPC does~not have such limitation. Conversely, the output-disturbance~DPC provides additional data in the form of disturbance estimates, which could be used, for example, for fault detection.


\section{Numerical comparison of formulations}
\label{section:Formulations comparison}

In this section, numerical examples are used to analyse and compare the velocity-form and the output-disturbance DPC in terms of their tracking and computational performances.

\subsection{Example 1: Velocity-Form DPC  Augmentation}
\label{subsection: Velocity-Form DPC}

This first example analyses the tracking performance of the velocity-form DPC \eqref{eq:7-DPC_entire_formulation_velocity}, but with boundary conditions enforced on trajectories of length $n+n_d$, rather than $n+p$. The aim is to examine the behaviour of the controller when the assumed model order, $n+n_d$, increases above $n+p$.  

For this comparison, we consider the problem of recovering tracking of the setpoint $y_{}^{s} = \begin{bmatrix} 0.65 & 0.77 \end{bmatrix}^\top$ after disturbances enter the plant \eqref{eq:plant}, with dynamics adapted from \cite{berberich_data-driven_2021-1},
\begin{gather*}
    \setlength{\abovedisplayskip}{5pt}
    \setlength{\belowdisplayskip}{0pt}
    A = \begin{bmatrix}
        0.921  & 0        & 0.041    & 0 \\ 
        0      & 0.918    & 0        & 0.033 \\
        0      & 0        & 0.924    & 0 \\
        0      & 0        & 0        & 0.937
    \end{bmatrix}, \quad 
    B = \begin{bmatrix}
        0.017  & 0.001 \\
        0.001  & 0.023 \\
        0      & 0.061 \\
        0.072  & 0
    \end{bmatrix}, \\
    C = \begin{bmatrix}
        1 & 0 & 0 & 0 \\
        0 & 1 & 0 & 0
    \end{bmatrix}, \quad
    (d_x,d_y) = \begin{cases}
        (0,0) & k \leq 100, \\
        (0.05,0.05) & k > 100.
    \end{cases} 
\end{gather*}
The DPC has prediction and data horizons $N = 20$ and $T = 107$; weights $Q = 10^2 \cdot \eye_p$, $R = 10^2 \cdot \eye_m$, $\lambda_{\alpha} = 10^{-2}$, and $\lambda_{\sigma} = 10^8$; and an input constraint set $\mathbb{U} = [-1, 1.5]^2$. The data $(\Delta u_{}^d,y_{}^d)$ are collected by applying to the system an input sequence $u_{}^d$, uniformly sampled from the interval $[0,1]^{T+1}$, and determining the corresponding $(\Delta u_{}^d,y_{}^d)$. $\Delta u_{}^d$ is PE of sufficient order for all simulated scenarios. The observer uses $\gamma=1$, corresponding to a deadbeat observer.

\begin{figure}[b] 
    \centering
    \input{Figures/velocityFormDPC_augmentation_comparison}
    \caption{Output trajectories for the velocity-form DPC in Section \ref{subsection: Velocity-Form DPC} with different value of assumed model order. Although transient behaviour varies, all formulations recover offset-free tracking.}
    \label{fig:velocityFormDPC_augmentation_comparison}
\end{figure}
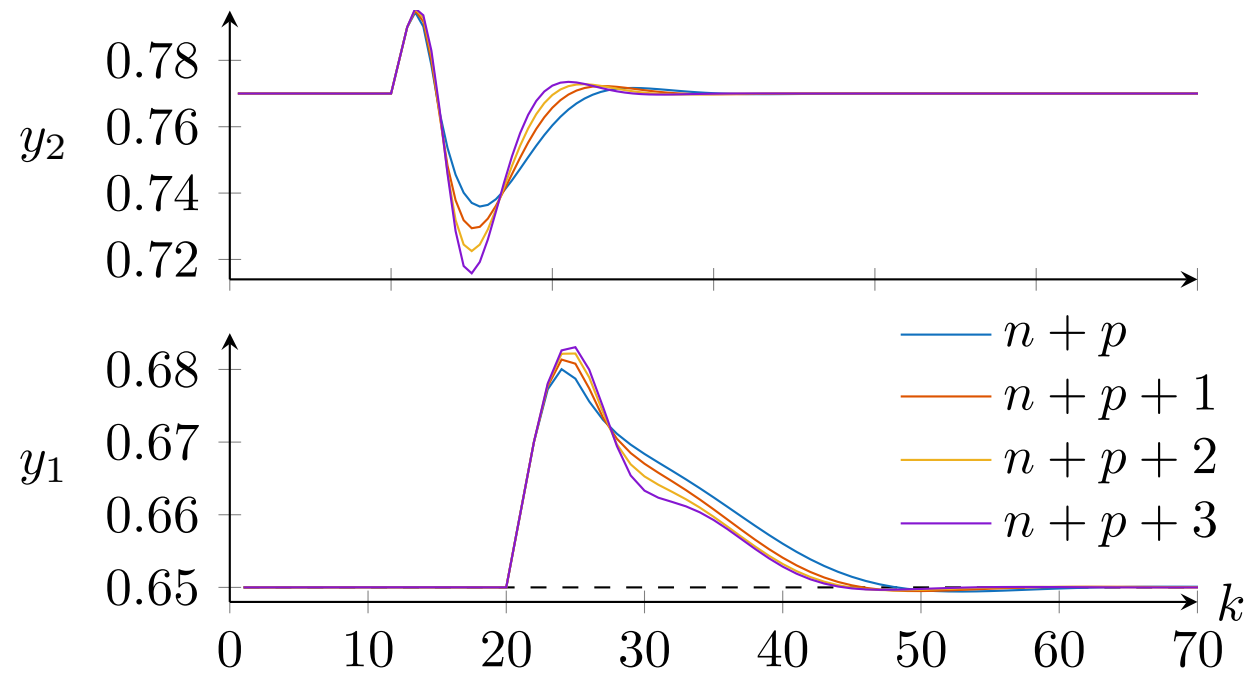

Fig. \ref{fig:velocityFormDPC_augmentation_comparison} shows the output response for different values of $n_d$. As expected from the discussions in Section \ref{section:velocity_formulation}, all model orders with $n_d \geq p$ correctly incorporate the velocity form and achieve offset-free tracking. It is interesting to notice that the different lengths of boundary conditions generate different transient behaviours, but investigating this phenomenon is beyond the scope of the present example.

\subsection{Example 2: Velocity-form vs. Output-Disturbance DPC}

This example compares the velocity-form and the output-disturbance DPC using the well-stirred chemical reactor from \cite[Example 1.11]{rawlings_model_2020}. Although this system violates the problem statement in Section \ref{section:problem_statement}, as the plant is nonlinear, is it used to show the offset-free tracking capabilities of the formulations.

For the simulations, $u_{}^d$, $\Delta u_{}^d$, and $y_{}^d$ are obtained from the linearised and discretised model 
\begin{align*}
    \setlength{\abovedisplayskip}{5pt}
    \setlength{\belowdisplayskip}{5pt}
    A &= \begin{bmatrix}
        0.2681  & -0.00338  & -0.00728  \\ 
        9.703   & 0.3279    & -25.44    \\
        0       & 0         & 1
    \end{bmatrix},  \\
    B &= \begin{bmatrix}
        -0.00537  & 0.1655 \\
        1.297  & 97.91 \\
        0      & -6.637
    \end{bmatrix}, \quad 
    C = \begin{bmatrix}
        1 & 0 & 0 \\
        0 & 0 & 1
    \end{bmatrix}.
\end{align*}
The matrix $C$ was modified from the original example such that only the first and third states are tracked. Moreover, as this model has a pole at $z=1$, it is expected that the output-disturbance DPC does not achieve offset-free tracking.

The DPCs were designed with an unconstrained input, $\mathbb{U} = \mathbb{R}^3$, and the following parameters: $N = 30$, $T = 125$, $Q = 10^3 \cdot \eye_p$, $R = \eye_m$, $\lambda_{\alpha} = 10^{-1}$, $\lambda_{\sigma} = 10^6$, and $\gamma=1$. The objective is to recover offset-free tracking of $y_{}^{s} = (0.878,0.659)$ after a $10\%$ increase in inlet flowrate at $k=10$. The reader may refer to \cite[Example 1.11]{rawlings_model_2020} for more details about the problem.

\begin{figure}[!hb] 
    \centering
    \input{Figures/integratingState_comparison}
    \caption{Output trajectories of the standard DPC \eqref{eq:data_driven_MPC_entire_formulation}, the velocity-form DPC \eqref{eq:7-DPC_entire_formulation_velocity}, and the output-disturbance DPC \eqref{eq:7-DPC_disturbance_mode_robust} for the model \emph{with} integrating modes in Section \ref{subsection: Velocity-Form DPC}. Only the velocity-form DPC achieves offset-free~tracking.}
    \label{fig:integratingState_comparison}
\end{figure}
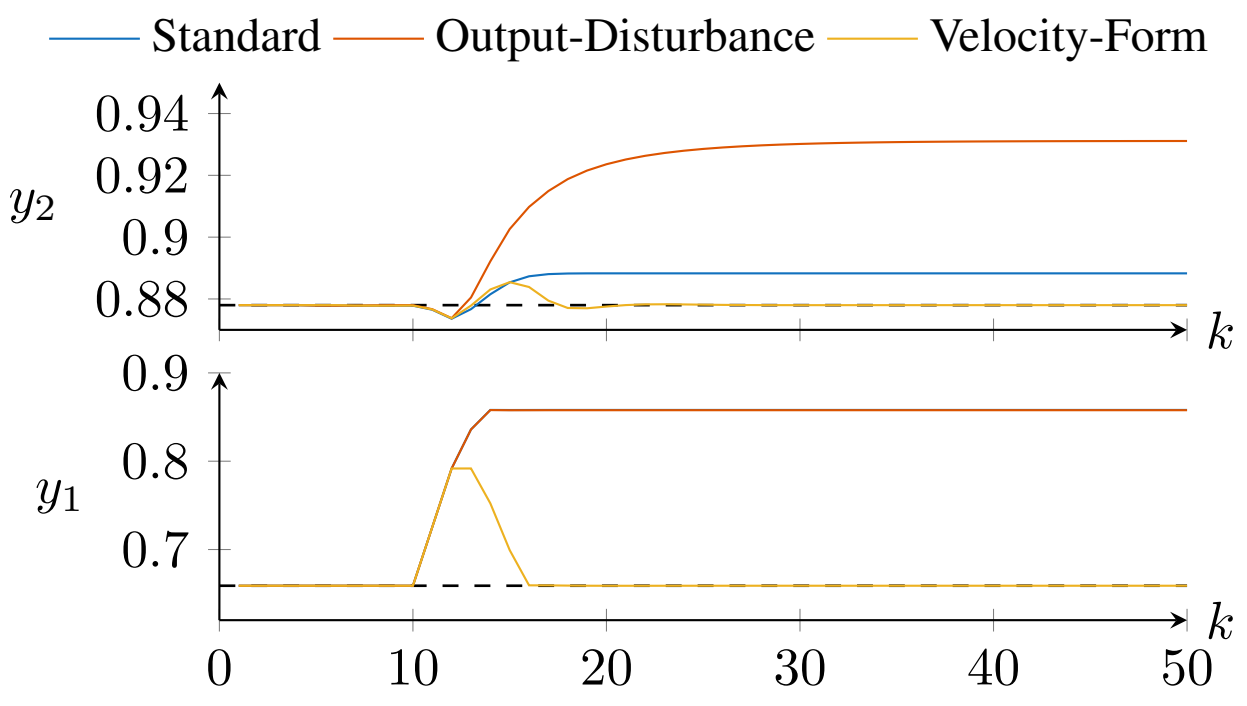

Fig. \ref{fig:integratingState_comparison} compares the tracking performance of the standard DPC \eqref{eq:data_driven_MPC_entire_formulation} with the velocity-form DPC \eqref{eq:7-DPC_entire_formulation_velocity} and the output-disturbance DPC \eqref{eq:7-DPC_disturbance_mode_robust}. As expected, the standard DPC does not recover offset-free tracking as it does not incorporate integral action. Similarly expected, the output-disturbance DPC does not recover offset tracking as the system contains a pole at $z=1$. The velocity-formulation DPC, however, successfully recovers tracking of the desired setpoint after the disturbance enters the system.

\begin{figure}[!hb] 
    \centering
    \input{Figures/integratingState_comparison2}
    \caption{Output trajectories of the standard DPC \eqref{eq:data_driven_MPC_entire_formulation}, the velocity-form DPC \eqref{eq:7-DPC_entire_formulation_velocity}, and the output-disturbance DPC \eqref{eq:7-DPC_disturbance_mode_robust} for the model \emph{without} integrating modes in Section \ref{subsection: Velocity-Form DPC}. Both the velocity-form DPC and the output-disturbance DPC display steady-state offset achieve offset-free tracking.}
    \label{fig:integratingState_comparison2}
\end{figure}
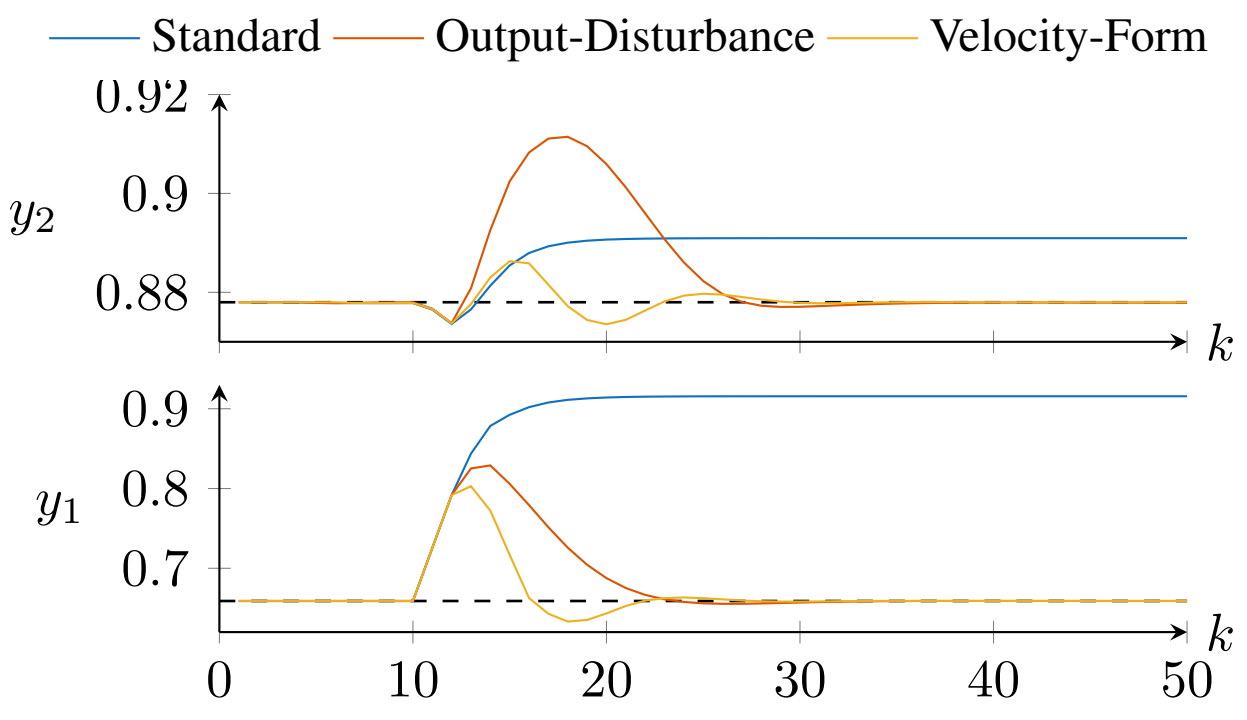

Fig. \ref{fig:integratingState_comparison2} compares the performances of the DPC with data $u_{}^d$, $\Delta u_{}^d$, and $y_{}^d$ coming from the system with 
\begin{equation*}
    A = \begin{bmatrix}
        0.2681  & -0.00338  & -0.00728  \\ 
        9.703   & 0.3279    & -25.44    \\
        0       & 0         & 0.9
    \end{bmatrix}
\end{equation*}
which does not contain the pole at $z=1$. For the same underlying nonlinear plant, the output-disturbance DPC now recovers offset-free tracking. Furthermore, the average iteration time over 10 simulations was $6.5$\,ms for the output-disturbance DPC and $7.5$\,ms for the velocity-form DPC, which shows that avoiding trajectory augmentation can result in improved computational performance. 

%% file: Figures/velocityFormDPC_augmentation_comparison.tex
%
%
\definecolor{mycolor1}{rgb}{0.06600,0.44300,0.74500}%
\definecolor{mycolor2}{rgb}{0.86600,0.32900,0.00000}%
\definecolor{mycolor3}{rgb}{0.92900,0.69400,0.12500}%
\definecolor{mycolor4}{rgb}{0.52100,0.08600,0.81900}%
\definecolor{mycolor5}{rgb}{0.23100,0.66600,0.19600}%
\begin{tikzpicture}

\begin{axis}[%
    width=0.9\figW,
    height=0.25\figH,
    at={(0\figW,0\figH)},
    scale only axis,
    xmin=0,
    xmax=70,
    ymin=0.648,
    ymax=0.685,
    axis background/.style={fill=white},
    axis lines = left,
    ylabel = $y_1$,
    xlabel = $k$,
    ylabel style = {rotate = -90},
    legend style={legend cell align=left, align=right, at={(1.05,1.15)}, draw=none},
    xlabel style={at={(current axis.south east)},anchor=west},
]
\addplot [dashed, forget plot]
  table[row sep=crcr]{%
1	0.77\\
200 0.77\\
};

\addplot [dashed, forget plot]
  table[row sep=crcr]{%
1	0.65\\
200 0.65\\
};

\addplot [color=mycolor1]
  table[row sep=crcr]{%
1	0.65\\
2	0.6500026\\
3	0.6500048429\\
4	0.6500067684401\\
5	0.650008412344713\\
6	0.650009806706396\\
7	0.650010980334301\\
8	0.650012106330534\\
9	0.650013020825986\\
10	0.650013634196974\\
11	0.65001391189265\\
12	0.650013858470712\\
13	0.650013504616073\\
14	0.650012896820515\\
15	0.650012089431954\\
16	0.650011138720215\\
17	0.650010098656405\\
18	0.650009018093903\\
19	0.650007939077121\\
20	0.650006896029909\\
21	0.66000591560573\\
22	0.670005017016606\\
23	0.677263469820782\\
24	0.680036670682022\\
25	0.678732230711483\\
26	0.675596451408698\\
27	0.673061301580066\\
28	0.671156191111728\\
29	0.66964975341294\\
30	0.668369785167277\\
31	0.667194307571556\\
32	0.666042936856634\\
33	0.664868777132217\\
34	0.663650936876235\\
35	0.662387806137229\\
36	0.661091134548413\\
37	0.659781001168816\\
38	0.65848168682417\\
39	0.657218422006253\\
40	0.656014953304296\\
41	0.654891848453754\\
42	0.653865444672778\\
43	0.652947336619703\\
44	0.65214429826678\\
45	0.651458536211891\\
46	0.650888179305495\\
47	0.650427919742383\\
48	0.650069732911666\\
49	0.649803616283819\\
50	0.649618300650997\\
51	0.649501899452085\\
52	0.649442473222748\\
53	0.649428496111176\\
54	0.649449219674629\\
55	0.649494935799724\\
56	0.649557145591374\\
57	0.649628644563403\\
58	0.649703536613309\\
59	0.649777190276308\\
60	0.649846150819878\\
61	0.649908021086837\\
62	0.649961322827881\\
63	0.650005348739877\\
64	0.650040013725753\\
65	0.650065712138929\\
66	0.65008318605204\\
67	0.650093408006747\\
68	0.650097480294609\\
69	0.650096551613066\\
70	0.650091750955078\\
71	0.650084137836791\\
72	0.650074667416197\\
73	0.65006416869287\\
74	0.650053333788431\\
75	0.65004271624154\\
76	0.650032736307783\\
77	0.650023691389925\\
78	0.650015769911927\\
79	0.650009067180209\\
80	0.650003602014488\\
81	0.649999333183456\\
82	0.649996174907577\\
83	0.649994010907262\\
84	0.649992706668589\\
85	0.64999211975632\\
86	0.649992108137148\\
87	0.649992536579718\\
88	0.649993281270604\\
89	0.6499942328456\\
90	0.64999529805524\\
91	0.649996400300637\\
92	0.649997479279862\\
93	0.649998489964379\\
94	0.64999940110381\\
95	0.650000193445602\\
96	0.650000857812526\\
97	0.650001393163705\\
98	0.650001804730379\\
99	0.650002102297222\\
100	0.65000229867542\\
101	0.650002408393243\\
102	0.650002446613588\\
103	0.650002428276821\\
104	0.650002367455584\\
105	0.650002276901104\\
106	0.650002167759032\\
107	0.650002049426225\\
108	0.650001929523812\\
109	0.650001813958277\\
110	0.65000170705004\\
111	0.65000161170354\\
112	0.650001529605199\\
113	0.650001461432058\\
114	0.650001407056543\\
115	0.650001365741807\\
116	0.650001336318972\\
117	0.650001317344426\\
118	0.650001307229836\\
119	0.650001304350507\\
120	0.650001307129823\\
121	0.650001314100732\\
122	0.650001323948054\\
123	0.650001335534183\\
124	0.650001347909492\\
125	0.650001360311901\\
126	0.650001372158364\\
127	0.650001383029409\\
128	0.650001392650688\\
129	0.65000140087081\\
130	0.650001407639228\\
131	0.65000141298454\\
132	0.650001416994836\\
133	0.650001419799366\\
134	0.65000142155348\\
135	0.650001422423594\\
136	0.650001422578799\\
137	0.650001422181155\\
138	0.650001421379597\\
139	0.650001420306928\\
140	0.650001419077475\\
141	0.650001417785771\\
142	0.65000141650579\\
143	0.65000141529411\\
144	0.650001414190959\\
145	0.650001413221848\\
146	0.650001412399742\\
147	0.650001411728344\\
148	0.650001411202572\\
149	0.650001410812895\\
150	0.65000141054497\\
151	0.650001410382648\\
152	0.650001410308088\\
153	0.650001410304105\\
154	0.650001410353678\\
155	0.650001410442785\\
156	0.650001410557241\\
157	0.650001410685928\\
158	0.65000141081946\\
159	0.650001410950315\\
160	0.650001411073317\\
161	0.650001411184421\\
162	0.650001411281375\\
163	0.650001411363046\\
164	0.650001411429217\\
165	0.650001411480262\\
166	0.650001411517666\\
167	0.650001411542363\\
168	0.650001411556828\\
169	0.650001411562719\\
170	0.650001411561969\\
171	0.650001411555912\\
172	0.650001411545948\\
173	0.650001411533258\\
174	0.650001411519607\\
175	0.650001411505747\\
176	0.650001411492061\\
177	0.650001411479399\\
178	0.650001411468348\\
179	0.650001411458994\\
180	0.650001411451411\\
181	0.650001411445094\\
182	0.650001411440466\\
183	0.65000141143738\\
184	0.650001411435421\\
185	0.65000141143419\\
186	0.650001411433929\\
187	0.650001411434261\\
188	0.65000141143465\\
189	0.650001411435398\\
190	0.650001411436071\\
191	0.650001411437002\\
192	0.650001411438315\\
193	0.650001411439579\\
194	0.650001411440741\\
195	0.650001411442146\\
196	0.65000141144305\\
197	0.65000141144375\\
198	0.650001411444538\\
199	0.650001411444812\\
200	0.65000141144522\\
};
\addlegendentry{$n+p$}

\addplot [color=mycolor2]
  table[row sep=crcr]{%
1	0.65\\
2	0.6500026\\
3	0.6500048429\\
4	0.6500067684401\\
5	0.650008412344713\\
6	0.650009806706396\\
7	0.650010980334301\\
8	0.650011959070416\\
9	0.650013091208145\\
10	0.650014034952245\\
11	0.650014603635342\\
12	0.650014723251504\\
13	0.65001439734161\\
14	0.650013678821762\\
15	0.65001264809579\\
16	0.650011396691329\\
17	0.65001001564475\\
18	0.650008587778115\\
19	0.650007183134098\\
20	0.650005856838012\\
21	0.660004648747221\\
22	0.670003584351639\\
23	0.67769294919013\\
24	0.681357418643622\\
25	0.680795545905489\\
26	0.67739039530573\\
27	0.673356384754005\\
28	0.670467832864345\\
29	0.668491246802414\\
30	0.667021223634666\\
31	0.665782322328733\\
32	0.664602157507456\\
33	0.663386906321485\\
34	0.662099823692775\\
35	0.660743015802241\\
36	0.659342518948898\\
37	0.657936598562264\\
38	0.656566953834572\\
39	0.655272569439076\\
40	0.654085822501972\\
41	0.653030424269173\\
42	0.652120781367411\\
43	0.65136238970319\\
44	0.650752917835126\\
45	0.650283689661098\\
46	0.649941333016771\\
47	0.649709416757223\\
48	0.649569950916302\\
49	0.649504670288897\\
50	0.649496060114856\\
51	0.649528112913611\\
52	0.649586828158108\\
53	0.649660482007591\\
54	0.649739703633669\\
55	0.649817398875073\\
56	0.649888562174057\\
57	0.649950015027169\\
58	0.650000104551943\\
59	0.6500383900339\\
60	0.650065339188903\\
61	0.650082049862526\\
62	0.650090007384502\\
63	0.650090883010708\\
64	0.650086374991739\\
65	0.650078090787165\\
66	0.650067466801771\\
67	0.650055720655264\\
68	0.650043830279379\\
69	0.650032533991189\\
70	0.650022345916815\\
71	0.650013581681677\\
72	0.650006389988027\\
73	0.650000786494686\\
74	0.649996687233865\\
75	0.649993939553299\\
76	0.649992349280777\\
77	0.649991703387218\\
78	0.649991787898636\\
79	0.649992401189437\\
80	0.649993363031649\\
81	0.649994519950652\\
82	0.64999574753223\\
83	0.649996950330135\\
84	0.649998060019329\\
85	0.64999903236768\\
86	0.649999843529448\\
87	0.650000486066749\\
88	0.650000965017479\\
89	0.650001294244238\\
90	0.650001493209529\\
91	0.650001584265906\\
92	0.650001590494909\\
93	0.650001534076425\\
94	0.650001435154073\\
95	0.650001311128109\\
96	0.650001176308937\\
97	0.650001041853316\\
98	0.650000915920379\\
99	0.650000803964698\\
100	0.650000709121865\\
101	0.650000632635004\\
102	0.650000574284527\\
103	0.650000532792853\\
104	0.650000506185618\\
105	0.650000492101294\\
106	0.65000048804039\\
107	0.650000491556398\\
108	0.650000500393457\\
109	0.650000512573697\\
110	0.650000526445751\\
111	0.650000540696971\\
112	0.650000554344876\\
113	0.650000566709009\\
114	0.650000577370278\\
115	0.650000586129314\\
116	0.650000592956785\\
117	0.650000597951275\\
118	0.650000601298018\\
119	0.650000603234232\\
120	0.650000604020397\\
121	0.650000603918215\\
122	0.650000603173391\\
123	0.650000602005421\\
124	0.650000600599414\\
125	0.650000599107563\\
126	0.650000597644087\\
127	0.650000596291632\\
128	0.65000059510419\\
129	0.650000594109461\\
130	0.650000593317335\\
131	0.650000592722455\\
132	0.650000592308752\\
133	0.650000592052109\\
134	0.65000059192581\\
135	0.650000591901828\\
136	0.650000591955017\\
137	0.650000592061334\\
138	0.650000592200041\\
139	0.650000592352893\\
140	0.650000592507269\\
141	0.650000592653245\\
142	0.650000592784051\\
143	0.650000592894581\\
144	0.650000592984223\\
145	0.650000593053513\\
146	0.650000593103589\\
147	0.650000593136773\\
148	0.650000593155325\\
149	0.650000593161881\\
150	0.650000593159341\\
151	0.650000593150643\\
152	0.650000593137197\\
153	0.650000593121867\\
154	0.650000593105677\\
155	0.650000593090122\\
156	0.65000059307615\\
157	0.650000593063968\\
158	0.650000593054303\\
159	0.65000059304643\\
160	0.650000593040413\\
161	0.650000593036252\\
162	0.650000593033648\\
163	0.650000593032509\\
164	0.650000593032358\\
165	0.650000593033014\\
166	0.650000593034207\\
167	0.650000593036116\\
168	0.650000593038031\\
169	0.650000593040074\\
170	0.650000593041203\\
171	0.65000059304229\\
172	0.650000593042869\\
173	0.650000593043056\\
174	0.65000059304369\\
175	0.650000593044251\\
176	0.650000593044042\\
177	0.650000593044212\\
178	0.650000593043804\\
179	0.650000593043657\\
180	0.650000593043772\\
181	0.650000593043474\\
182	0.650000593043331\\
183	0.650000593042648\\
184	0.650000593041838\\
185	0.650000593041918\\
186	0.65000059304218\\
187	0.650000593042886\\
188	0.650000593043215\\
189	0.650000593043318\\
190	0.65000059304354\\
191	0.650000593043822\\
192	0.650000593043797\\
193	0.650000593044019\\
194	0.650000593044101\\
195	0.650000593044922\\
196	0.650000593044839\\
197	0.650000593045083\\
198	0.650000593045425\\
199	0.650000593045322\\
200	0.650000593045035\\
};
\addlegendentry{$n+p+1$}

\addplot [color=mycolor3]
  table[row sep=crcr]{%
1	0.65\\
2	0.6500026\\
3	0.6500048429\\
4	0.6500067684401\\
5	0.650008412344713\\
6	0.650009806706396\\
7	0.650010980334301\\
8	0.650011959070416\\
9	0.650012766076505\\
10	0.650014010964723\\
11	0.650015067431478\\
12	0.650015608416574\\
13	0.650015516971782\\
14	0.650014813920422\\
15	0.650013601356595\\
16	0.650012020374203\\
17	0.650010221148606\\
18	0.650008343184371\\
19	0.650006503762007\\
20	0.650004792600773\\
21	0.660003270977096\\
22	0.670001973828119\\
23	0.677935609235722\\
24	0.682158958848626\\
25	0.682194698534674\\
26	0.678909804090768\\
27	0.674095247421112\\
28	0.669706249604616\\
29	0.666941529477215\\
30	0.66527611014854\\
31	0.6641318795642\\
32	0.663148597289682\\
33	0.662128715198614\\
34	0.660989640622725\\
35	0.659724230121861\\
36	0.658369562554161\\
37	0.656983494201872\\
38	0.65562811140116\\
39	0.654359029730639\\
40	0.653219351052385\\
41	0.652237228545765\\
42	0.65142599534401\\
43	0.650785942905513\\
44	0.65030700250312\\
45	0.649971755834118\\
46	0.649758364086383\\
47	0.649643149109876\\
48	0.649602680205305\\
49	0.649615313489216\\
50	0.649662198608911\\
51	0.649727812229125\\
52	0.649800102856361\\
53	0.649870341237763\\
54	0.649932768866947\\
55	0.649984127758908\\
56	0.650023140853342\\
57	0.650049996726273\\
58	0.650065876669039\\
59	0.6500725479651\\
60	0.650072035138511\\
61	0.650066371444981\\
62	0.65005742594464\\
63	0.650046796973213\\
64	0.650035760345535\\
65	0.650025259808634\\
66	0.650015927670237\\
67	0.650008124758526\\
68	0.650001990616355\\
69	0.649997496753132\\
70	0.649994497707879\\
71	0.64999277643886\\
72	0.649992082066529\\
73	0.649992159207044\\
74	0.649992769041263\\
75	0.649993702888609\\
76	0.649994789424931\\
77	0.649995896857103\\
78	0.649996931372097\\
79	0.649997833087517\\
80	0.649998570563734\\
81	0.649999134733201\\
82	0.649999532896837\\
83	0.649999783237585\\
84	0.649999910124775\\
85	0.649999940342878\\
86	0.649999900260931\\
87	0.649999813885845\\
88	0.649999701684733\\
89	0.649999580038421\\
90	0.64999946117849\\
91	0.649999353471596\\
92	0.649999261921059\\
93	0.649999188780357\\
94	0.649999134201874\\
95	0.649999096850472\\
96	0.649999074454816\\
97	0.64999906426142\\
98	0.649999063396261\\
99	0.649999069123152\\
100	0.649999079015829\\
101	0.649999091052692\\
102	0.649999103653242\\
103	0.649999115668537\\
104	0.649999126340108\\
105	0.649999135242764\\
106	0.649999142216328\\
107	0.649999147298583\\
108	0.649999150663641\\
109	0.649999152566867\\
110	0.649999153299923\\
111	0.649999153157351\\
112	0.649999152414302\\
113	0.649999151308871\\
114	0.649999150036455\\
115	0.649999148745827\\
116	0.649999147542077\\
117	0.649999146492774\\
118	0.649999145632595\\
119	0.64999914497119\\
120	0.649999144499672\\
121	0.649999144197268\\
122	0.649999144036545\\
123	0.649999143987642\\
124	0.649999144020081\\
125	0.649999144107018\\
126	0.649999144225393\\
127	0.649999144356773\\
128	0.649999144486698\\
129	0.649999144605738\\
130	0.649999144708114\\
131	0.649999144790821\\
132	0.649999144852927\\
133	0.649999144896297\\
134	0.649999144922617\\
135	0.649999144935252\\
136	0.649999144937728\\
137	0.649999144932726\\
138	0.649999144922673\\
139	0.649999144909958\\
140	0.649999144896395\\
141	0.649999144883334\\
142	0.649999144871615\\
143	0.649999144861681\\
144	0.649999144853642\\
145	0.649999144848084\\
146	0.649999144844547\\
147	0.649999144842706\\
148	0.649999144841757\\
149	0.649999144841101\\
150	0.64999914484146\\
151	0.649999144842674\\
152	0.649999144843726\\
153	0.649999144844783\\
154	0.649999144845664\\
155	0.649999144846794\\
156	0.649999144848386\\
157	0.649999144849835\\
158	0.649999144850816\\
159	0.649999144851196\\
160	0.649999144851448\\
161	0.649999144851795\\
162	0.64999914485183\\
163	0.649999144851645\\
164	0.649999144851243\\
165	0.649999144851112\\
166	0.64999914485112\\
167	0.649999144851455\\
168	0.649999144851463\\
169	0.6499991448514\\
170	0.649999144851249\\
171	0.649999144850886\\
172	0.649999144850706\\
173	0.649999144850583\\
174	0.649999144850302\\
175	0.649999144850362\\
176	0.649999144850149\\
177	0.649999144849482\\
178	0.649999144848948\\
179	0.649999144848743\\
180	0.649999144848701\\
181	0.649999144848837\\
182	0.649999144848738\\
183	0.649999144848552\\
184	0.649999144848247\\
185	0.649999144848091\\
186	0.64999914484836\\
187	0.649999144849001\\
188	0.649999144849706\\
189	0.649999144849997\\
190	0.649999144850299\\
191	0.649999144850081\\
192	0.649999144850084\\
193	0.64999914485013\\
194	0.649999144850546\\
195	0.649999144850566\\
196	0.649999144850664\\
197	0.64999914485063\\
198	0.649999144850688\\
199	0.649999144850561\\
200	0.649999144850553\\
};
\addlegendentry{$n+p+2$}

\addplot [color=mycolor4]
  table[row sep=crcr]{%
1	0.65\\
2	0.6500026\\
3	0.6500048429\\
4	0.6500067684401\\
5	0.650008412344713\\
6	0.650009806706396\\
7	0.650010980334301\\
8	0.650011959070416\\
9	0.650012766076505\\
10	0.650013422094392\\
11	0.650014995807347\\
12	0.650016393789419\\
13	0.650017069112152\\
14	0.65001684651249\\
15	0.650015782823411\\
16	0.65001406089007\\
17	0.650011913333219\\
18	0.650009571711808\\
19	0.650007236028977\\
20	0.650005059697099\\
21	0.660003145723547\\
22	0.670001550094345\\
23	0.6780696695002\\
24	0.682623677054547\\
25	0.683065361935563\\
26	0.679977607658751\\
27	0.674834259674614\\
28	0.669461434647544\\
29	0.665407359274733\\
30	0.66331520487909\\
31	0.662346666803262\\
32	0.661763685545069\\
33	0.661154792266025\\
34	0.66033714842078\\
35	0.659275848649123\\
36	0.658021041745858\\
37	0.656661807151469\\
38	0.65529478687712\\
39	0.654004951992025\\
40	0.652855747133595\\
41	0.651886044499217\\
42	0.651111575688756\\
43	0.650528975714553\\
44	0.650120960638443\\
45	0.649861587300681\\
46	0.649720927151893\\
47	0.649668798575406\\
48	0.649677437953033\\
49	0.649723152786167\\
50	0.649787099870762\\
51	0.649855380166925\\
52	0.649918652692486\\
53	0.649971454802258\\
54	0.650011386194051\\
55	0.650038277339882\\
56	0.650053425916583\\
57	0.650058951292586\\
58	0.650057289499556\\
59	0.650050830258777\\
60	0.650041683418964\\
61	0.650031553859054\\
62	0.650021700370178\\
63	0.650012954053796\\
64	0.650005774170414\\
65	0.650000323125635\\
66	0.649996546574025\\
67	0.649994248847032\\
68	0.64999315768231\\
69	0.64999297532911\\
70	0.64999341544919\\
71	0.649994226841846\\
72	0.649995205998982\\
73	0.649996200928793\\
74	0.649997108749203\\
75	0.649997869334095\\
76	0.649998456932768\\
77	0.649998871239198\\
78	0.649999128961567\\
79	0.649999256541535\\
80	0.64999928434825\\
81	0.649999242419425\\
82	0.64999915765003\\
83	0.649999052219942\\
84	0.649998943001761\\
85	0.649998841680423\\
86	0.64999875533515\\
87	0.649998687277354\\
88	0.64999863797447\\
89	0.649998605943671\\
90	0.649998588537768\\
91	0.649998582583664\\
92	0.649998584860761\\
93	0.649998592426489\\
94	0.649998602804853\\
95	0.649998614070612\\
96	0.649998624851559\\
97	0.649998634275975\\
98	0.649998641890966\\
99	0.649998647566546\\
100	0.649998651401699\\
101	0.649998653639292\\
102	0.649998654593778\\
103	0.649998654598569\\
104	0.649998653968124\\
105	0.649998652972829\\
106	0.649998651827978\\
107	0.649998650691886\\
108	0.649998649669734\\
109	0.649998648821639\\
110	0.649998648170429\\
111	0.64999864771319\\
112	0.649998647429182\\
113	0.649998647287851\\
114	0.649998647255341\\
115	0.649998647298709\\
116	0.649998647388402\\
117	0.649998647500832\\
118	0.649998647617973\\
119	0.649998647726728\\
120	0.64999864781961\\
121	0.649998647893085\\
122	0.649998647946629\\
123	0.649998647981432\\
124	0.649998648000677\\
125	0.649998648007701\\
126	0.649998648005861\\
127	0.649998647998424\\
128	0.649998647987625\\
129	0.649998647975668\\
130	0.649998647964379\\
131	0.649998647954465\\
132	0.649998647946336\\
133	0.649998647940404\\
134	0.649998647936384\\
135	0.649998647934028\\
136	0.649998647932813\\
137	0.649998647932574\\
138	0.649998647932862\\
139	0.649998647933529\\
140	0.649998647934678\\
141	0.649998647935724\\
142	0.649998647936606\\
143	0.649998647937456\\
144	0.649998647938191\\
145	0.649998647938742\\
146	0.649998647939115\\
147	0.649998647939423\\
148	0.649998647939407\\
149	0.649998647939504\\
150	0.649998647939782\\
151	0.649998647939876\\
152	0.649998647940022\\
153	0.649998647939881\\
154	0.649998647939402\\
155	0.649998647939041\\
156	0.649998647938641\\
157	0.649998647938362\\
158	0.649998647938264\\
159	0.649998647938129\\
160	0.6499986479381\\
161	0.649998647938387\\
162	0.649998647938491\\
163	0.649998647938367\\
164	0.649998647938469\\
165	0.649998647938639\\
166	0.649998647938608\\
167	0.649998647938663\\
168	0.649998647938714\\
169	0.649998647938898\\
170	0.649998647939109\\
171	0.649998647939546\\
172	0.649998647939792\\
173	0.649998647939693\\
174	0.649998647939515\\
175	0.649998647939095\\
176	0.649998647938901\\
177	0.649998647938566\\
178	0.649998647938874\\
179	0.649998647938861\\
180	0.64999864793902\\
181	0.649998647939315\\
182	0.649998647939518\\
183	0.649998647939724\\
184	0.64999864793981\\
185	0.649998647939786\\
186	0.649998647939589\\
187	0.649998647939483\\
188	0.649998647939654\\
189	0.649998647939916\\
190	0.649998647939791\\
191	0.649998647939556\\
192	0.649998647939318\\
193	0.649998647939217\\
194	0.649998647939321\\
195	0.649998647939438\\
196	0.649998647939541\\
197	0.649998647939671\\
198	0.649998647939835\\
199	0.649998647939645\\
200	0.649998647939484\\
};
\addlegendentry{$n + p+3$}
\end{axis}

\begin{axis}[%
    width=0.9\figW,
    height=0.25\figH,
    at={(0\figW,0.3\figH)},
    scale only axis,
    xmin=0,
    xmax=120,
    ymin=0.714,
    ymax=0.795,
    axis background/.style={fill=white},
    axis lines = left,
    xticklabels=\empty,
    ylabel = $y_2$,
    ylabel style = {rotate = -90},
    legend style={legend cell align=left, align=right, at={(1,0.7)}}
]

\addplot [color=mycolor1, forget plot]
  table[row sep=crcr]{%
1	0.77\\
2	0.7700009\\
3	0.7700016668\\
4	0.7700023150646\\
5	0.770002858020144\\
6	0.770003307587511\\
7	0.770003674503077\\
8	0.77000270207229\\
9	0.770000966462237\\
10	0.769998904281745\\
11	0.769996831037747\\
12	0.769994960234653\\
13	0.769993422136691\\
14	0.769992281412138\\
15	0.769991553144725\\
16	0.769991216877184\\
17	0.769991228538275\\
18	0.769991530247301\\
19	0.769992058070849\\
20	0.769992747904083\\
21	0.779993539677837\\
22	0.789994380140904\\
23	0.794452688467465\\
24	0.790185625229448\\
25	0.778529645380507\\
26	0.76456864589518\\
27	0.753610408400064\\
28	0.745534748630583\\
29	0.740119007911012\\
30	0.737045933657008\\
31	0.735947704790776\\
32	0.736440285805644\\
33	0.73814907517562\\
34	0.740726892702967\\
35	0.743865556696137\\
36	0.747302256885834\\
37	0.750821793569035\\
38	0.754255696580983\\
39	0.75747912146298\\
40	0.760406287184093\\
41	0.762985084982782\\
42	0.76519135845026\\
43	0.767023235903853\\
44	0.768495790364547\\
45	0.769636211734744\\
46	0.770479600439651\\
47	0.771065431368732\\
48	0.771434690420597\\
49	0.771627651788134\\
50	0.771682240716304\\
51	0.771632912023564\\
52	0.771509967522246\\
53	0.771339233985439\\
54	0.771142026059819\\
55	0.770935324222\\
56	0.770732105480919\\
57	0.770541773170293\\
58	0.770370641128489\\
59	0.770222436355517\\
60	0.770098792468938\\
61	0.769999713696339\\
62	0.769923995626239\\
63	0.769869594405539\\
64	0.769833940535342\\
65	0.769814196918238\\
66	0.769807463463223\\
67	0.769810932392714\\
68	0.769821999613969\\
69	0.769838338161738\\
70	0.769857939941011\\
71	0.769879131859294\\
72	0.769900572094702\\
73	0.769921231676409\\
74	0.769940365951478\\
75	0.769957479796912\\
76	0.769972289775134\\
77	0.769984685748089\\
78	0.769994693875706\\
79	0.770002442369934\\
80	0.770008130904838\\
81	0.770012004201792\\
82	0.770014329975303\\
83	0.770015381187928\\
84	0.770015422372712\\
85	0.770014699658934\\
86	0.770013434051135\\
87	0.770011817479256\\
88	0.770010011126477\\
89	0.770008145558498\\
90	0.770006322208671\\
91	0.770004615829972\\
92	0.770003077559828\\
93	0.770001738318736\\
94	0.770000612305366\\
95	0.769999700401985\\
96	0.769998993367251\\
97	0.769998474713114\\
98	0.769998123218736\\
99	0.769997915050058\\
100	0.769997825490288\\
101	0.769997830289149\\
102	0.769997906666778\\
103	0.769998034007677\\
104	0.769998194289059\\
105	0.769998372287354\\
106	0.769998555605462\\
107	0.769998734565067\\
108	0.769998901999889\\
109	0.769999052985018\\
110	0.769999184524869\\
111	0.76999929523055\\
112	0.76999938500032\\
113	0.769999454716233\\
114	0.769999505968622\\
115	0.769999540815297\\
116	0.769999561575396\\
117	0.769999570661184\\
118	0.769999570443696\\
119	0.769999563156125\\
120	0.769999550823338\\
121	0.769999535218148\\
122	0.769999517842332\\
123	0.76999949991977\\
124	0.769999482407527\\
125	0.769999466015496\\
126	0.769999451233444\\
127	0.769999438360515\\
128	0.769999427534952\\
129	0.769999418769586\\
130	0.769999411978285\\
131	0.769999407004005\\
132	0.769999403644345\\
133	0.769999401670116\\
134	0.76999940084239\\
135	0.769999400927592\\
136	0.769999401704301\\
137	0.769999402972343\\
138	0.769999404556801\\
139	0.7699994063092\\
140	0.7699994081085\\
141	0.769999409860551\\
142	0.769999411496193\\
143	0.769999412968112\\
144	0.769999414246485\\
145	0.769999415318574\\
146	0.7699994161841\\
147	0.76999941685137\\
148	0.769999417337648\\
149	0.769999417663531\\
150	0.769999417851773\\
151	0.769999417926038\\
152	0.769999417909972\\
153	0.769999417825581\\
154	0.769999417693552\\
155	0.769999417531192\\
156	0.769999417353671\\
157	0.769999417172994\\
158	0.76999941699762\\
159	0.769999416834719\\
160	0.769999416687988\\
161	0.769999416561027\\
162	0.769999416454828\\
163	0.76999941636943\\
164	0.769999416303356\\
165	0.769999416255587\\
166	0.769999416224322\\
167	0.769999416206547\\
168	0.769999416199529\\
169	0.769999416201667\\
170	0.769999416210238\\
171	0.769999416223288\\
172	0.769999416239985\\
173	0.769999416257902\\
174	0.769999416276825\\
175	0.769999416295204\\
176	0.769999416312594\\
177	0.769999416328276\\
178	0.769999416341653\\
179	0.769999416352253\\
180	0.769999416360463\\
181	0.769999416366582\\
182	0.769999416371154\\
183	0.769999416373732\\
184	0.769999416375782\\
185	0.769999416377018\\
186	0.769999416377032\\
187	0.769999416375909\\
188	0.769999416373869\\
189	0.76999941637145\\
190	0.769999416369398\\
191	0.769999416367204\\
192	0.769999416365262\\
193	0.769999416363592\\
194	0.769999416361244\\
195	0.769999416358851\\
196	0.769999416357429\\
197	0.769999416356905\\
198	0.769999416357094\\
199	0.769999416356494\\
200	0.769999416356395\\
};

\addplot [color=mycolor2, forget plot]
  table[row sep=crcr]{%
1	0.77\\
2	0.7700009\\
3	0.7700016668\\
4	0.7700023150646\\
5	0.770002858020144\\
6	0.770003307587511\\
7	0.770003674503077\\
8	0.770003968428889\\
9	0.770002265478384\\
10	0.769999586302662\\
11	0.769996654188106\\
12	0.769993946563267\\
13	0.769991745139307\\
14	0.769990182378845\\
15	0.769989282673218\\
16	0.769988997444726\\
17	0.769989233837225\\
18	0.769989877157235\\
19	0.769990807626469\\
20	0.769991912080676\\
21	0.779993091407597\\
22	0.789994264544035\\
23	0.795029979537378\\
24	0.791695485630747\\
25	0.780063720184789\\
26	0.763760664377344\\
27	0.748253051774173\\
28	0.737847828308469\\
29	0.731821517464249\\
30	0.7294118298125\\
31	0.729825258429196\\
32	0.732300951623829\\
33	0.736152989451163\\
34	0.740794905289826\\
35	0.745750139280869\\
36	0.750652060214751\\
37	0.755236691387771\\
38	0.759330827602621\\
39	0.762837661785043\\
40	0.765721588137718\\
41	0.767993414129762\\
42	0.769696820839123\\
43	0.770896582047741\\
44	0.771668788745527\\
45	0.772093124527358\\
46	0.772247092871855\\
47	0.772202001724933\\
48	0.772020455754434\\
49	0.771755083516977\\
50	0.771448227720091\\
51	0.771132344521493\\
52	0.770830886433582\\
53	0.770559477810137\\
54	0.77032722835889\\
55	0.770138065647788\\
56	0.769992000322384\\
57	0.769886266405261\\
58	0.769816303071908\\
59	0.769776563523943\\
60	0.769761151179782\\
61	0.769764293843715\\
62	0.76978067332107\\
63	0.769805631751492\\
64	0.769835277363539\\
65	0.769866511994623\\
66	0.769897001110016\\
67	0.769925104644177\\
68	0.769949784139381\\
69	0.76997049871222\\
70	0.769987099459761\\
71	0.769999729297881\\
72	0.77000873288735\\
73	0.770014579361399\\
74	0.770017799019215\\
75	0.770018933950113\\
76	0.770018501717251\\
77	0.770016970663899\\
78	0.770014745113063\\
79	0.770012158591989\\
80	0.770009473255219\\
81	0.770006883821685\\
82	0.770004524518746\\
83	0.770002477772623\\
84	0.770000783638179\\
85	0.76999944917954\\
86	0.769998457248314\\
87	0.769997774297511\\
88	0.769997357017922\\
89	0.769997157719376\\
90	0.769997128486056\\
91	0.769997224190837\\
92	0.769997404492954\\
93	0.769997634995618\\
94	0.769997887714204\\
95	0.7699981410204\\
96	0.769998379213959\\
97	0.769998591850835\\
98	0.769998772929114\\
99	0.769998920036486\\
100	0.769999033506999\\
101	0.769999115645469\\
102	0.769999170040349\\
103	0.769999200993959\\
104	0.769999213061071\\
105	0.769999210702005\\
106	0.769999198032707\\
107	0.76999917867106\\
108	0.769999155649184\\
109	0.769999131391579\\
110	0.769999107730821\\
111	0.769999085959907\\
112	0.769999066901039\\
113	0.769999050982622\\
114	0.769999038323704\\
115	0.769999028810133\\
116	0.769999022165393\\
117	0.769999018012595\\
118	0.769999015924646\\
119	0.76999901546409\\
120	0.769999016214057\\
121	0.769999017798095\\
122	0.769999019889536\\
123	0.769999022221205\\
124	0.769999024585036\\
125	0.769999026825836\\
126	0.769999028839\\
127	0.769999030563056\\
128	0.769999031972011\\
129	0.769999033064322\\
130	0.769999033859917\\
131	0.769999034391552\\
132	0.769999034698545\\
133	0.769999034822795\\
134	0.769999034806759\\
135	0.769999034689165\\
136	0.769999034505398\\
137	0.769999034285101\\
138	0.769999034051608\\
139	0.7699990338215\\
140	0.769999033608242\\
141	0.769999033421637\\
142	0.769999033265928\\
143	0.769999033142255\\
144	0.769999033049575\\
145	0.769999032984553\\
146	0.769999032944228\\
147	0.76999903292382\\
148	0.769999032919386\\
149	0.769999032927068\\
150	0.769999032942312\\
151	0.769999032963606\\
152	0.769999032986797\\
153	0.769999033009486\\
154	0.769999033030266\\
155	0.769999033049212\\
156	0.769999033066069\\
157	0.769999033081129\\
158	0.769999033092437\\
159	0.769999033100849\\
160	0.769999033106535\\
161	0.769999033110528\\
162	0.769999033111278\\
163	0.769999033109615\\
164	0.76999903310624\\
165	0.769999033102687\\
166	0.769999033099615\\
167	0.769999033096668\\
168	0.769999033094104\\
169	0.769999033092152\\
170	0.769999033091863\\
171	0.769999033092451\\
172	0.769999033093318\\
173	0.769999033093108\\
174	0.769999033092737\\
175	0.769999033092618\\
176	0.769999033093069\\
177	0.769999033093209\\
178	0.769999033093605\\
179	0.769999033093325\\
180	0.769999033092485\\
181	0.769999033091489\\
182	0.769999033090593\\
183	0.769999033090231\\
184	0.769999033090639\\
185	0.769999033090514\\
186	0.769999033090188\\
187	0.769999033089918\\
188	0.76999903309043\\
189	0.769999033091228\\
190	0.76999903309168\\
191	0.769999033093124\\
192	0.769999033093235\\
193	0.76999903309236\\
194	0.769999033091963\\
195	0.769999033091218\\
196	0.769999033090642\\
197	0.769999033090301\\
198	0.769999033089908\\
199	0.769999033090137\\
200	0.769999033090206\\
};

\addplot [color=mycolor3, forget plot]
  table[row sep=crcr]{%
1	0.77\\
2	0.7700009\\
3	0.7700016668\\
4	0.7700023150646\\
5	0.770002858020144\\
6	0.770003307587511\\
7	0.770003674503077\\
8	0.770003968428889\\
9	0.770004198052976\\
10	0.770001431251406\\
11	0.769997432428376\\
12	0.769993368803602\\
13	0.769989936716226\\
14	0.769987478819188\\
15	0.769986087280552\\
16	0.769985690268405\\
17	0.769986120996726\\
18	0.769987170063518\\
19	0.769988622579591\\
20	0.769990282509619\\
21	0.779991986590894\\
22	0.789993610257239\\
23	0.795416800642376\\
24	0.792847912478387\\
25	0.781704398744935\\
26	0.764578054540083\\
27	0.746257536790744\\
28	0.731869875676353\\
29	0.724495364815166\\
30	0.722523598962701\\
31	0.724480733351002\\
32	0.729038029072471\\
33	0.735073294910217\\
34	0.741695532815875\\
35	0.748242933684677\\
36	0.754263025217309\\
37	0.759482444191993\\
38	0.763772169868254\\
39	0.767112391365302\\
40	0.769559983832563\\
41	0.771220265958088\\
42	0.772223892070107\\
43	0.772709058016149\\
44	0.772808723149402\\
45	0.77264225184262\\
46	0.772310722573973\\
47	0.77189510633775\\
48	0.771456547253198\\
49	0.771038059249644\\
50	0.770667061236473\\
51	0.770358291856693\\
52	0.770116761294693\\
53	0.769940503381444\\
54	0.769822981585375\\
55	0.769755075385872\\
56	0.769726628881604\\
57	0.769727582686791\\
58	0.769748735295966\\
59	0.769782193715068\\
60	0.769821577904397\\
61	0.769862041873086\\
62	0.769900168312672\\
63	0.769933785251196\\
64	0.7699617437616\\
65	0.769983686342037\\
66	0.769999826881943\\
67	0.770010755597447\\
68	0.770017276134268\\
69	0.770020277298852\\
70	0.770020638408276\\
71	0.770019164992788\\
72	0.770016550263342\\
73	0.7700133572356\\
74	0.770010016409179\\
75	0.770006834308902\\
76	0.770004008840824\\
77	0.770001648151893\\
78	0.769999790462843\\
79	0.769998423047154\\
80	0.769997499177626\\
81	0.769996952376371\\
82	0.769996707724568\\
83	0.769996690281667\\
84	0.769996830876932\\
85	0.769997069646086\\
86	0.769997357751169\\
87	0.769997657713385\\
88	0.769997942769289\\
89	0.76999819560208\\
90	0.769998406743064\\
91	0.769998572865714\\
92	0.769998695135925\\
93	0.769998777731085\\
94	0.769998826585424\\
95	0.769998848389599\\
96	0.769998849843449\\
97	0.769998837139444\\
98	0.769998815642922\\
99	0.76999878973462\\
100	0.769998762779226\\
101	0.769998737183356\\
102	0.769998714503993\\
103	0.769998695591096\\
104	0.769998680742856\\
105	0.769998669851626\\
106	0.769998662537788\\
107	0.769998658263542\\
108	0.769998656422535\\
109	0.769998656406946\\
110	0.769998657655826\\
111	0.769998659684064\\
112	0.769998662095525\\
113	0.769998664586007\\
114	0.769998666938397\\
115	0.76999866901347\\
116	0.769998670737292\\
117	0.769998672085337\\
118	0.76999867306796\\
119	0.769998673721025\\
120	0.769998674096283\\
121	0.769998674250056\\
122	0.769998674238355\\
123	0.769998674113362\\
124	0.769998673919357\\
125	0.769998673693087\\
126	0.769998673461742\\
127	0.769998673244278\\
128	0.769998673053178\\
129	0.769998672894926\\
130	0.769998672771682\\
131	0.769998672682525\\
132	0.769998672624686\\
133	0.769998672591958\\
134	0.769998672579136\\
135	0.769998672581978\\
136	0.769998672594506\\
137	0.769998672612072\\
138	0.769998672633104\\
139	0.769998672655136\\
140	0.769998672676006\\
141	0.769998672694711\\
142	0.769998672710079\\
143	0.769998672720999\\
144	0.769998672729159\\
145	0.769998672734266\\
146	0.769998672736887\\
147	0.769998672737519\\
148	0.769998672736818\\
149	0.76999867273515\\
150	0.769998672732965\\
151	0.769998672730327\\
152	0.769998672728299\\
153	0.769998672726248\\
154	0.76999867272498\\
155	0.769998672723771\\
156	0.769998672721957\\
157	0.769998672721564\\
158	0.769998672721683\\
159	0.76999867272227\\
160	0.76999867272287\\
161	0.769998672722875\\
162	0.769998672722691\\
163	0.769998672723575\\
164	0.769998672724147\\
165	0.769998672724408\\
166	0.769998672724657\\
167	0.769998672724363\\
168	0.769998672723644\\
169	0.769998672722242\\
170	0.769998672721493\\
171	0.769998672720812\\
172	0.769998672720523\\
173	0.769998672720311\\
174	0.769998672720414\\
175	0.769998672720885\\
176	0.769998672721435\\
177	0.769998672722065\\
178	0.769998672722345\\
179	0.769998672722753\\
180	0.769998672723235\\
181	0.769998672723248\\
182	0.769998672723327\\
183	0.76999867272327\\
184	0.76999867272289\\
185	0.769998672722512\\
186	0.769998672721908\\
187	0.769998672721673\\
188	0.769998672721292\\
189	0.769998672721493\\
190	0.769998672721656\\
191	0.769998672722638\\
192	0.769998672723508\\
193	0.769998672723765\\
194	0.769998672723182\\
195	0.769998672722584\\
196	0.769998672722687\\
197	0.769998672723366\\
198	0.769998672724196\\
199	0.76999867272455\\
200	0.76999867272482\\
};

\addplot [color=mycolor4, forget plot]
  table[row sep=crcr]{%
1	0.77\\
2	0.7700009\\
3	0.7700016668\\
4	0.7700023150646\\
5	0.770002858020144\\
6	0.770003307587511\\
7	0.770003674503077\\
8	0.770003968428889\\
9	0.770004198052976\\
10	0.770004371180666\\
11	0.770000348983657\\
12	0.769994934176979\\
13	0.769989849311721\\
14	0.769986004491446\\
15	0.769983736184961\\
16	0.769983007793066\\
17	0.769983568554481\\
18	0.769985071828404\\
19	0.769987156849576\\
20	0.769989499908975\\
21	0.779991841397464\\
22	0.789993995111623\\
23	0.79564601236644\\
24	0.793585170580833\\
25	0.782914945655313\\
26	0.765605465226712\\
27	0.745815503213231\\
28	0.728556847539876\\
29	0.71802444598323\\
30	0.715822142568598\\
31	0.719231400381325\\
32	0.72601741012058\\
33	0.734399458171242\\
34	0.743061979215612\\
35	0.751117335430189\\
36	0.758040888600808\\
37	0.763594265872863\\
38	0.76774829733937\\
39	0.770613111653074\\
40	0.77237949415571\\
41	0.773273069033106\\
42	0.773521224547345\\
43	0.77333138005113\\
44	0.77287867618287\\
45	0.772300963239433\\
46	0.771699015740475\\
47	0.771140132044299\\
48	0.770663601997518\\
49	0.770286879511122\\
50	0.770011636671329\\
51	0.769829174455811\\
52	0.769724908595242\\
53	0.769681833979618\\
54	0.769683000348417\\
55	0.7697131131244\\
56	0.769759415377891\\
57	0.769812019739586\\
58	0.769863851624612\\
59	0.769910345180451\\
60	0.769949007160392\\
61	0.769978936076089\\
62	0.770000357666715\\
63	0.77001421489847\\
64	0.77002183223031\\
65	0.770024660029063\\
66	0.770024095421166\\
67	0.770021369964937\\
68	0.770017491589944\\
69	0.770013227497255\\
70	0.77000911548026\\
71	0.77000549279717\\
72	0.770002533842048\\
73	0.770000290054819\\
74	0.76999872756776\\
75	0.769997759853192\\
76	0.769997274040638\\
77	0.769997150625558\\
78	0.769997277008003\\
79	0.769997555715378\\
80	0.769997908374436\\
81	0.769998276523159\\
82	0.769998620273733\\
83	0.769998915689567\\
84	0.769999151567101\\
85	0.769999326126085\\
86	0.769999443951304\\
87	0.769999513384566\\
88	0.769999544456248\\
89	0.769999547367976\\
90	0.769999531477115\\
91	0.769999504708589\\
92	0.769999473301554\\
93	0.769999441799341\\
94	0.769999413201586\\
95	0.769999389211446\\
96	0.769999370517277\\
97	0.769999357074642\\
98	0.769999348360679\\
99	0.769999343585819\\
100	0.769999341859346\\
101	0.769999342306023\\
102	0.769999344144886\\
103	0.76999934672957\\
104	0.769999349563219\\
105	0.769999352294458\\
106	0.769999354701639\\
107	0.769999356667885\\
108	0.769999358157601\\
109	0.76999935919269\\
110	0.769999359829638\\
111	0.769999360143586\\
112	0.769999360214338\\
113	0.769999360117624\\
114	0.769999359919179\\
115	0.76999935967296\\
116	0.769999359418596\\
117	0.769999359182278\\
118	0.76999935897975\\
119	0.76999935881926\\
120	0.769999358700996\\
121	0.769999358622324\\
122	0.769999358576947\\
123	0.769999358558117\\
124	0.769999358558856\\
125	0.769999358572275\\
126	0.769999358593216\\
127	0.769999358616835\\
128	0.76999935864006\\
129	0.769999358660641\\
130	0.769999358677352\\
131	0.769999358689972\\
132	0.769999358699078\\
133	0.769999358705037\\
134	0.769999358708131\\
135	0.769999358709114\\
136	0.769999358707904\\
137	0.769999358706125\\
138	0.769999358704061\\
139	0.769999358701684\\
140	0.769999358700298\\
141	0.769999358698682\\
142	0.769999358697417\\
143	0.769999358696736\\
144	0.769999358695942\\
145	0.769999358695806\\
146	0.769999358695617\\
147	0.769999358695903\\
148	0.769999358696169\\
149	0.769999358696418\\
150	0.769999358696101\\
151	0.769999358695805\\
152	0.769999358695207\\
153	0.769999358695442\\
154	0.769999358696015\\
155	0.769999358696047\\
156	0.769999358696179\\
157	0.769999358696676\\
158	0.769999358696686\\
159	0.769999358696199\\
160	0.76999935869633\\
161	0.769999358695917\\
162	0.769999358695377\\
163	0.769999358695068\\
164	0.769999358694629\\
165	0.769999358695043\\
166	0.769999358695358\\
167	0.769999358695753\\
168	0.7699993586959\\
169	0.769999358695988\\
170	0.769999358696052\\
171	0.769999358696083\\
172	0.76999935869581\\
173	0.769999358696041\\
174	0.769999358696315\\
175	0.769999358696346\\
176	0.769999358696254\\
177	0.76999935869654\\
178	0.769999358696731\\
179	0.769999358697152\\
180	0.76999935869712\\
181	0.769999358696926\\
182	0.769999358696385\\
183	0.769999358696152\\
184	0.769999358695926\\
185	0.769999358695769\\
186	0.769999358695711\\
187	0.76999935869564\\
188	0.769999358695612\\
189	0.769999358695149\\
190	0.769999358695214\\
191	0.769999358695531\\
192	0.769999358695864\\
193	0.76999935869625\\
194	0.769999358696611\\
195	0.769999358696961\\
196	0.769999358697527\\
197	0.76999935869739\\
198	0.769999358696739\\
199	0.769999358696561\\
200	0.769999358696633\\
};

\end{axis}
\end{tikzpicture}%

%% file: Figures/integratingState_comparison.tex
%
%
\definecolor{mycolor1}{rgb}{0.06600,0.44300,0.74500}%
\definecolor{mycolor2}{rgb}{0.86600,0.32900,0.00000}%
\definecolor{mycolor3}{rgb}{0.92900,0.69400,0.12500}%
\begin{tikzpicture}

\begin{axis}[%
    width=0.9\figW,
    height=0.23\figH,
    at={(0\figW,0.27\figH)},
    scale only axis,
    xmin=0,
    xmax=50,
    ymin=0.87,
    ymax=0.95,
    axis background/.style={fill=white},
    axis lines = left,
    legend columns=-1,
    ylabel = $y_2$,
    ylabel style = {rotate = -90},
    legend style={font=\small, legend cell align=left, align=left, draw=none, at={(1.05,1.35)}},
    xticklabels=\empty,
    xlabel = $k$,
    xlabel style={at={(current axis.south east)},anchor=west},
]

\addplot [dashed, forget plot, line width=0.5pt]
  table[row sep=crcr]{%
0 0.878     \\
50 0.878    \\
};

\addplot [color=mycolor1]
  table[row sep=crcr]{%
1	0.878\\
2	0.878\\
3	0.878\\
4	0.878\\
5	0.877860593361282\\
6	0.877825188293553\\
7	0.877821027423506\\
8	0.877823383056015\\
9	0.87782530958972\\
10	0.877826060354678\\
11	0.876511742910995\\
12	0.873602809186063\\
13	0.876697929368903\\
14	0.881517534593866\\
15	0.885370662868856\\
16	0.887337886119146\\
17	0.888041178975277\\
18	0.888239953660269\\
19	0.888283301509474\\
20	0.888288593705422\\
21	0.888287504634221\\
22	0.888286449979391\\
23	0.888285987623239\\
24	0.888285836906672\\
25	0.888285798508401\\
26	0.888285791711708\\
27	0.888285791573517\\
28	0.888285792117809\\
29	0.888285792421169\\
30	0.888285792533784\\
31	0.88828579256639\\
32	0.888285792573565\\
33	0.888285792574441\\
34	0.888285792574247\\
35	0.888285792574062\\
36	0.88828579257398\\
37	0.888285792573947\\
38	0.888285792573959\\
39	0.888285792573983\\
40	0.888285792574003\\
41	0.88828579257401\\
42	0.888285792573996\\
43	0.888285792573982\\
44	0.888285792573965\\
45	0.88828579257397\\
46	0.888285792573959\\
47	0.888285792573959\\
48	0.888285792573939\\
49	0.888285792573955\\
50	0.888285792573955\\
};
\addlegendentry{Standard}

\addplot [color=mycolor2]
  table[row sep=crcr]{%
1	0.878\\
2	0.878\\
3	0.878\\
4	0.878\\
5	0.877860593361282\\
6	0.877825188293553\\
7	0.877871913807174\\
8	0.877945847755035\\
9	0.877986385511652\\
10	0.877997733649686\\
11	0.876687477374974\\
12	0.873786287000591\\
13	0.880467684768278\\
14	0.892172151908229\\
15	0.902608710085671\\
16	0.909786213486132\\
17	0.914924607765918\\
18	0.918777920335856\\
19	0.921556609849193\\
20	0.923589073434623\\
21	0.925133416311906\\
22	0.9263184549753\\
23	0.927238547111439\\
24	0.927965974220512\\
25	0.928546683083259\\
26	0.92901337912171\\
27	0.92939128665438\\
28	0.929698916556379\\
29	0.929950240327941\\
30	0.930156252465125\\
31	0.930325571223018\\
32	0.930465003413635\\
33	0.930580014366645\\
34	0.930675011098879\\
35	0.930753560054951\\
36	0.930818565852267\\
37	0.93087240254486\\
38	0.93091701507306\\
39	0.930954001268174\\
40	0.930984676580824\\
41	0.931010125592594\\
42	0.931031243798904\\
43	0.931048771440273\\
44	0.931063320974272\\
45	0.931075399514523\\
46	0.931085427232911\\
47	0.931093752446953\\
48	0.931100664028422\\
49	0.931106401649767\\
50	0.931111164203044\\
};
\addlegendentry{Output-Disturbance}

\addplot [color=mycolor3]
  table[row sep=crcr]{%
1	0.878\\
2	0.878\\
3	0.878\\
4	0.878\\
5	0.878\\
6	0.878\\
7	0.877860593361282\\
8	0.877825188293553\\
9	0.877832337818007\\
10	0.877866247888862\\
11	0.876602183205685\\
12	0.873737447217513\\
13	0.8778496692324\\
14	0.883072745375999\\
15	0.885431311237924\\
16	0.883870148000237\\
17	0.879454616571216\\
18	0.877081870800064\\
19	0.876988874337316\\
20	0.877551394858771\\
21	0.878020092330864\\
22	0.878246262023394\\
23	0.878283580469715\\
24	0.87822786707897\\
25	0.878150260293551\\
26	0.878084580468185\\
27	0.878039863355627\\
28	0.878013913285976\\
29	0.878001206894356\\
30	0.877996420701546\\
31	0.877995649287423\\
32	0.877996489059763\\
33	0.877997682100677\\
34	0.877998695106771\\
35	0.877999385867764\\
36	0.877999786459822\\
37	0.87799998233827\\
38	0.878000055961274\\
39	0.878000067648065\\
40	0.878000054496628\\
41	0.878000035939405\\
42	0.878000020213777\\
43	0.878000009503162\\
44	0.878000003297213\\
45	0.878000000265491\\
46	0.877999999127973\\
47	0.877999998949298\\
48	0.877999999154571\\
49	0.877999999442897\\
50	0.877999999686897\\
};
\addlegendentry{Velocity-Form}

\end{axis}

\begin{axis}[%
    width=0.9\figW,
    height=0.23\figH,
    at={(0\figW,0\figH)},
    scale only axis,
    xmin=0,
    xmax=50,
    ymin=0.62,
    ymax=0.9,
    axis lines = left,
    ylabel = $y_1$,
    ylabel style = {rotate = -90},
    xlabel = $k$,
    xlabel style={at={(current axis.south east)},anchor=west},
]

\addplot [dashed, forget plot, line width=0.5pt]
  table[row sep=crcr]{%
0 0.659     \\
50 0.659    \\
};

\addplot [color=mycolor1, forget plot]
  table[row sep=crcr]{%
1	0.659\\
2	0.659\\
3	0.659\\
4	0.659\\
5	0.659\\
6	0.659\\
7	0.659002680031085\\
8	0.659001572229998\\
9	0.65900155888307\\
10	0.659001564378881\\
11	0.72537005292143\\
12	0.791738538782048\\
13	0.835936581212936\\
14	0.858049012419838\\
15	0.857952867316635\\
16	0.857988629060437\\
17	0.858024839693979\\
18	0.85803291746244\\
19	0.858032661112477\\
20	0.858032386463701\\
21	0.858032263498202\\
22	0.858032217487145\\
23	0.858032203957741\\
24	0.858032201098154\\
25	0.858032200789103\\
26	0.858032200880966\\
27	0.858032200958159\\
28	0.858032200991003\\
29	0.858032201001521\\
30	0.858032201004163\\
31	0.858032201004592\\
32	0.858032201004604\\
33	0.858032201004576\\
34	0.858032201004542\\
35	0.858032201004533\\
36	0.858032201004533\\
37	0.858032201004536\\
38	0.858032201004534\\
39	0.858032201004528\\
40	0.858032201004539\\
41	0.85803220100453\\
42	0.858032201004532\\
43	0.858032201004521\\
44	0.85803220100454\\
45	0.858032201004526\\
46	0.85803220100453\\
47	0.858032201004534\\
48	0.858032201004529\\
49	0.85803220100453\\
50	0.858032201004528\\
};
\addplot [color=mycolor2, forget plot]
  table[row sep=crcr]{%
1	0.659\\
2	0.659\\
3	0.659\\
4	0.659\\
5	0.659\\
6	0.659\\
7	0.65900211786745\\
8	0.659000150008256\\
9	0.658999973297157\\
10	0.659000154068685\\
11	0.725368743520278\\
12	0.791737260641013\\
13	0.835888599828513\\
14	0.857886888686136\\
15	0.857697987437041\\
16	0.857719274044062\\
17	0.857756199906071\\
18	0.85775066422239\\
19	0.857749827870102\\
20	0.857755968571602\\
21	0.857757617372857\\
22	0.857758333366718\\
23	0.85775983451025\\
24	0.857760826079205\\
25	0.857761453197667\\
26	0.857762069202814\\
27	0.857762573106165\\
28	0.857762957413738\\
29	0.857763283030377\\
30	0.857763555186137\\
31	0.857763776465898\\
32	0.857763960221204\\
33	0.857764113446016\\
34	0.857764240356769\\
35	0.857764345884115\\
36	0.857764433895002\\
37	0.857764507267387\\
38	0.857764568523138\\
39	0.857764619761493\\
40	0.857764662682734\\
41	0.857764698702739\\
42	0.857764728997752\\
43	0.857764754539174\\
44	0.857764776128073\\
45	0.857764794443307\\
46	0.857764810036411\\
47	0.857764823366023\\
48	0.857764834813086\\
49	0.857764844705035\\
50	0.857764853298061\\
};
\addplot [color=mycolor3, forget plot]
  table[row sep=crcr]{%
1	0.659\\
2	0.659\\
3	0.659\\
4	0.659\\
5	0.659\\
6	0.659\\
7	0.659\\
8	0.659\\
9	0.659007324678338\\
10	0.659008084360738\\
11	0.725371794000117\\
12	0.79173736321431\\
13	0.791781320790141\\
14	0.75228573307602\\
15	0.699330514355199\\
16	0.659455875032209\\
17	0.659389487913768\\
18	0.659158627923416\\
19	0.659007116589061\\
20	0.658956116151395\\
21	0.658943656307439\\
22	0.658948974856448\\
23	0.65896566584908\\
24	0.658981681196348\\
25	0.658992273487156\\
26	0.658998093886393\\
27	0.659000807200073\\
28	0.659001586666558\\
29	0.659001438896523\\
30	0.659001019595328\\
31	0.659000613937574\\
32	0.659000315033528\\
33	0.659000128843152\\
34	0.659000030110707\\
35	0.658999987626571\\
36	0.658999975817904\\
37	0.658999977824235\\
38	0.658999984191548\\
39	0.658999990475996\\
40	0.658999995126526\\
41	0.658999998012015\\
42	0.658999999537683\\
43	0.659000000194045\\
44	0.659000000376101\\
45	0.659000000344298\\
46	0.659000000245149\\
47	0.659000000147561\\
48	0.659000000075446\\
49	0.659000000030735\\
50	0.659000000007111\\
};
\end{axis}
\end{tikzpicture}%

%% file: Figures/integratingState_comparison2.tex
%
%
\definecolor{mycolor1}{rgb}{0.06600,0.44300,0.74500}%
\definecolor{mycolor2}{rgb}{0.86600,0.32900,0.00000}%
\definecolor{mycolor3}{rgb}{0.92900,0.69400,0.12500}%
\begin{tikzpicture}

\begin{axis}[%
    width=0.9\figW,
    height=0.23\figH,
    at={(0\figW,0.27\figH)},
    scale only axis,
    xmin=0,
    xmax=50,
    ymin=0.87,
    ymax=0.92,
    axis lines = left,
    legend columns=-1,
    ylabel = $y_2$,
    ylabel style = {rotate = -90},
    legend style={font=\small, legend cell align=left, align=left, draw=none, at={(1.05,1.4)}},
    xticklabels=\empty,
    xlabel = $k$,
    xlabel style={at={(current axis.south east)},anchor=west},
]

\addplot [dashed, forget plot, line width=0.5pt]
  table[row sep=crcr]{%
0 0.878     \\
50 0.878    \\
};

\addplot [color=mycolor1]
  table[row sep=crcr]{%
1	0.878\\
2	0.878\\
3	0.878\\
4	0.878\\
5	0.877860593361282\\
6	0.877825188293553\\
7	0.877821035897958\\
8	0.877823336525874\\
9	0.877825243093638\\
10	0.877826012745027\\
11	0.876511718829984\\
12	0.873602799078594\\
13	0.876553024509194\\
14	0.881351291579109\\
15	0.885440234346301\\
16	0.887961062544728\\
17	0.889313752486253\\
18	0.890043604747413\\
19	0.890445052019793\\
20	0.890665735539567\\
21	0.89078891239009\\
22	0.890859258540429\\
23	0.890899752063758\\
24	0.890923104550829\\
25	0.890936633585972\\
26	0.890944495556123\\
27	0.890949064220951\\
28	0.890951719476774\\
29	0.890953263835053\\
30	0.890954162243422\\
31	0.890954684806691\\
32	0.890954988782129\\
33	0.890955165625513\\
34	0.890955268505777\\
35	0.890955328356074\\
36	0.890955363174765\\
37	0.890955383431272\\
38	0.890955395215782\\
39	0.890955402071591\\
40	0.890955406060061\\
41	0.890955408380435\\
42	0.890955409730369\\
43	0.890955410515696\\
44	0.890955410972595\\
45	0.890955411238419\\
46	0.890955411393029\\
47	0.890955411482998\\
48	0.89095541153535\\
49	0.890955411565776\\
50	0.890955411583475\\
};
\addlegendentry{Standard}

\addplot [color=mycolor2]
  table[row sep=crcr]{%
1	0.878\\
2	0.878\\
3	0.878\\
4	0.878\\
5	0.877860593361282\\
6	0.877825188293553\\
7	0.877872210966878\\
8	0.877946509428209\\
9	0.877987267350845\\
10	0.877998739762495\\
11	0.876688557155856\\
12	0.873787423156145\\
13	0.880826750351617\\
14	0.892597582264636\\
15	0.90245391101475\\
16	0.908272913725112\\
17	0.911090412295682\\
18	0.91145318620329\\
19	0.909547968408025\\
20	0.905927194492927\\
21	0.901219801268778\\
22	0.895972246405754\\
23	0.890739007128566\\
24	0.886048507458771\\
25	0.882292726708639\\
26	0.879648756809885\\
27	0.878059638830332\\
28	0.877292023159375\\
29	0.877049818888491\\
30	0.877078914267207\\
31	0.877214270968291\\
32	0.877371738543972\\
33	0.877517035963051\\
34	0.877639019273365\\
35	0.877735441049111\\
36	0.87780757784143\\
37	0.877858466843619\\
38	0.877892009787065\\
39	0.877912263780251\\
40	0.877922939411084\\
41	0.877927141011096\\
42	0.877927294744573\\
43	0.877925187021463\\
44	0.877922055485557\\
45	0.877918698295678\\
46	0.877915582359159\\
47	0.877912939735508\\
48	0.8779108470128\\
49	0.877909286421287\\
50	0.877908190012462\\
};
\addlegendentry{Output-Disturbance}

\addplot [color=mycolor3]
  table[row sep=crcr]{%
1	0.878\\
2	0.878\\
3	0.878\\
4	0.878\\
5	0.878\\
6	0.878\\
7	0.877860593361282\\
8	0.877825188293553\\
9	0.877832003646228\\
10	0.877864997578174\\
11	0.876599711085575\\
12	0.873734066222909\\
13	0.877618388160864\\
14	0.883025030818312\\
15	0.886298465411803\\
16	0.885867534337949\\
17	0.881548994060182\\
18	0.877196172943074\\
19	0.874413571827279\\
20	0.873543080361301\\
21	0.874443429314454\\
22	0.876321426505669\\
23	0.878170465646082\\
24	0.879332320794186\\
25	0.879719852950727\\
26	0.879533802103099\\
27	0.879057084372376\\
28	0.878536927741744\\
29	0.878127251519069\\
30	0.877884478165992\\
31	0.877794480615659\\
32	0.877809625739274\\
33	0.877875404958117\\
34	0.877948126645384\\
35	0.878002642780404\\
36	0.878031171857953\\
37	0.878037369007544\\
38	0.878029779771788\\
39	0.878017043428894\\
40	0.878005291473031\\
41	0.87799750174953\\
42	0.877994097922271\\
43	0.87799402246649\\
44	0.877995738587005\\
45	0.877997902861523\\
46	0.877999669440162\\
47	0.878000705726435\\
48	0.878001049458856\\
49	0.878000924043214\\
50	0.878000586029236\\
};
\addlegendentry{Velocity-Form}

\end{axis}

\begin{axis}[%
    width=0.9\figW,
    height=0.23\figH,
    at={(0\figW,0\figH)},
    scale only axis,
    xmin=0,
    xmax=50,
    ymin=0.62,
    ymax=0.93,
    axis lines = left,
    ylabel = $y_1$,
    ylabel style = {rotate = -90},
    xlabel = $k$,
    xlabel style={at={(current axis.south east)},anchor=west},
]

\addplot [dashed, forget plot, line width=0.5pt]
  table[row sep=crcr]{%
0 0.659     \\
50 0.659    \\
};

\addplot [color=mycolor1, forget plot]
  table[row sep=crcr]{%
1	0.659\\
2	0.659\\
3	0.659\\
4	0.659\\
5	0.659\\
6	0.659\\
7	0.659002142729917\\
8	0.659001429191809\\
9	0.659001537804283\\
10	0.659001617160896\\
11	0.725370144265144\\
12	0.791738626625043\\
13	0.843763143608119\\
14	0.878475659210878\\
15	0.892423247753668\\
16	0.902008448118658\\
17	0.907845665060051\\
18	0.9111540335621\\
19	0.912994394297827\\
20	0.914109360407706\\
21	0.914757251402564\\
22	0.915130313137313\\
23	0.915347063544727\\
24	0.915474083341458\\
25	0.915547785136821\\
26	0.915590592932738\\
27	0.915615520613863\\
28	0.915630034807475\\
29	0.915638471872401\\
30	0.91564338004659\\
31	0.915646236274297\\
32	0.915647897941584\\
33	0.915648864498723\\
34	0.915649426839095\\
35	0.915649754004987\\
36	0.915649944334721\\
37	0.915650055060536\\
38	0.915650119478374\\
39	0.915650156954675\\
40	0.915650178757016\\
41	0.915650191440936\\
42	0.915650198820056\\
43	0.915650203112978\\
44	0.915650205610464\\
45	0.915650207063427\\
46	0.915650207908709\\
47	0.915650208400472\\
48	0.915650208686565\\
49	0.915650208852999\\
50	0.915650208949834\\
};
\addplot [color=mycolor2, forget plot]
  table[row sep=crcr]{%
1	0.659\\
2	0.659\\
3	0.659\\
4	0.659\\
5	0.659\\
6	0.659\\
7	0.659001779801292\\
8	0.658999932536375\\
9	0.658999833350568\\
10	0.658999945165178\\
11	0.725368493636659\\
12	0.791736996789599\\
13	0.82513619715771\\
14	0.82895665966703\\
15	0.805778319196312\\
16	0.779173398972203\\
17	0.751442404386135\\
18	0.725883250676543\\
19	0.704394878789067\\
20	0.687647164557393\\
21	0.675262930353724\\
22	0.666650749396041\\
23	0.661068240506006\\
24	0.657767486453311\\
25	0.656083969719101\\
26	0.6554763583884\\
27	0.655527691012225\\
28	0.655936238774555\\
29	0.656497159042167\\
30	0.65708117620687\\
31	0.65761442516077\\
32	0.658061550287527\\
33	0.658412337598539\\
34	0.658671634361553\\
35	0.658852106546427\\
36	0.658969357865434\\
37	0.659038906351015\\
38	0.659074526539433\\
39	0.659087526529728\\
40	0.659086618601568\\
41	0.659078131012449\\
42	0.659066385028036\\
43	0.65905412212792\\
44	0.659042912324559\\
45	0.659033507487406\\
46	0.659026125718458\\
47	0.659020666619208\\
48	0.659016864901968\\
49	0.659014393198553\\
50	0.659012925596138\\
};
\addplot [color=mycolor3, forget plot]
  table[row sep=crcr]{%
1	0.659\\
2	0.659\\
3	0.659\\
4	0.659\\
5	0.659\\
6	0.659\\
7	0.659\\
8	0.659\\
9	0.659007342405716\\
10	0.659009293775304\\
11	0.725374189104037\\
12	0.791739357752147\\
13	0.802870836279088\\
14	0.77228905485523\\
15	0.716804141487584\\
16	0.662645391459942\\
17	0.642993600406524\\
18	0.633236147002004\\
19	0.635233041534104\\
20	0.643457904646021\\
21	0.652737485227179\\
22	0.659436419798846\\
23	0.662611000899798\\
24	0.663428075759839\\
25	0.662538797599144\\
26	0.661010423229529\\
27	0.659604551538416\\
28	0.658689965725322\\
29	0.658296847124999\\
30	0.658287742890068\\
31	0.658494698910798\\
32	0.658753616668736\\
33	0.658963327792971\\
34	0.65908526886161\\
35	0.659125223214589\\
36	0.659109822970539\\
37	0.659069251343003\\
38	0.659027362241075\\
39	0.658997080302085\\
40	0.658981841740945\\
41	0.658979064149558\\
42	0.658983737107129\\
43	0.658991001860248\\
44	0.658997504444329\\
45	0.659001693565829\\
46	0.659003423451841\\
47	0.659003337834495\\
48	0.659002311823142\\
49	0.659001087375187\\
50	0.659000117219043\\
};
\end{axis}
\end{tikzpicture}%